\documentclass[twocolumn]{aastex7}

\usepackage{amsthm,latexsym,amssymb,amsmath,amsfonts}
\usepackage{xcolor}
\usepackage{graphicx}
\usepackage{longtable}

\newcommand{\Drizzlepac}{\texttt{DrizzlePac 3.0}}
\newcommand{\TweakReg}{\texttt{TweakReg}}
\newcommand{\Astrodrizzle}{\texttt{AstroDrizzle}}
\newcommand{\Dolphot}{\texttt{DOLPHOT}}
\newcommand{\hstphot}{\texttt{HSTPhot}}

\newcommand{\msol}{M_{\odot}}

\newcommand{\ho}{\ensuremath{H_0}}
\newcommand{\degree}{$^\circ$}

\newcommand{\jwst}{\emph{JWST}\,}
\newcommand{\hst}{\emph{HST}\,}

\begin{document}

\title{Tip of the Red Giant Branch Distances to NGC~1316, NGC~1380, NGC~1404, \& NGC~4457: \\ A Pilot Study of a Parallel Distance Ladder Using Type Ia Supernovae in Early-Type Host Galaxies}

\author[0000-0002-8092-2077]{Max J. B. Newman}
\affiliation{Rutgers University, Department of Physics and Astronomy, 136 Frelinghuysen Road Piscataway, NJ 08854, USA}
\email[show]{mjbnewman25astro@gmail.com}

\author[0000-0003-2037-4619]{Conor Larison}
\affiliation{Rutgers University, Department of Physics and Astronomy, 136 Frelinghuysen Road Piscataway, NJ 08854, USA}
\email{cl449@physics.rutgers.edu}

\author[0000-0001-8738-6011]{Saurabh W. Jha}
\affiliation{Rutgers University, Department of Physics and Astronomy, 136 Frelinghuysen Road Piscataway, NJ 08854, USA}
\email{saurabh@physics.rutgers.edu}

\author[0000-0001-5538-2614]{Kristen B. W. McQuinn}
\affiliation{Rutgers University, Department of Physics and Astronomy, 136 Frelinghuysen Road Piscataway, NJ 08854, USA}
\affiliation{Space Telescope Science Institute, 3700 San Martin Drive, Baltimore, MD 21218, USA}
\email{kmcquinn@stsci.edu}

\author[0000-0003-0605-8732]{Evan D. Skillman}
\affiliation{University of Minnesota, Minnesota Institute for Astrophysics, School of Physics and Astronomy, 116 Church Street, S.E., Minneapolis,\\
MN 55455, USA}
\email{skill001@umn.edu}

\author[0000-0001-8416-4093]{Andrew E. Dolphin}
\affiliation{Raytheon Technologies, 1151 E. Hermans Road, Tucson, AZ 85756, USA}
\affiliation{Steward Observatory, University of Arizona, 933 North Cherry Avenue, Tucson, AZ 85721, USA}
\email{andy@dolphinism.com}

\author[0000-0002-5995-9692]{Mi Dai}
\affiliation{Department of Physics and Astronomy, University of Pittsburgh, 100 Allen Hall, 3941 O'Hara St, Pittsburgh, PA, 15260}
\email{mi.dai@pitt.edu}

\author[0000-0003-4253-656X]{D. Andrew Howell}
\affiliation{Las Cumbres Observatory, 6740 Cortona Dr., Suite 102, Goleta, CA, 93117, USA}
\affiliation{Department of Physics, University of California, Santa Barbara, Santa Barbara, CA, 93106, USA}
\email{dahowell@gmail.com}

\author[0000-0001-5807-7893]{Curtis McCully}
\affiliation{Las Cumbres Observatory, 6740 Cortona Dr., Suite 102, Goleta, CA, 93117, USA}
\affiliation{Department of Physics, University of California, Santa Barbara, Santa Barbara, CA, 93106, USA}
\email{cmccully@lco.global}

\author[0000-0002-4924-444X]{K. Azalee Bostroem}
\altaffiliation{LSST-DA Catalyst Fellow}
\affiliation{Steward Observatory, University of Arizona, 933 North Cherry Avenue, Tuscon, AZ 85721-0065, USA}
\email{bostroem@arizona.edu}

\author[0000-0002-1125-9187]{Daichi Hiramatsu}
\affiliation{Center for Astrophysics, Harvard \& Smithsonian, 60 Garden Street,
Cambridge, MA 02138-1516, USA}
\affiliation{The NSF AI Institute for Artificial Intelligence and Fundamental
Interactions, USA}
\email{daichi.hiramatsu@cfa.harvard.edu}


\author[0000-0002-7472-1279]{Craig Pellegrino}
\affiliation{Goddard Space Flight Center, 8800 Greenbelt Rd, Greenbelt, MD 20771, USA}
\email{craig.m.pellegrino@nasa.gov}

\author[0000-0003-0209-9246]{Estefania Padilla Gonzalez}
\affiliation{Las Cumbres Observatory, 6740 Cortona Dr., Suite 102, Goleta, CA, 93117, USA}
\affiliation{Department of Physics, University of California, Santa Barbara, Santa Barbara, CA, 93106, USA}
\email{epadill7@jh.edu}

\submitjournal{ApJ}
\received{August 29$^{\text{th}}$, 2025}
\begin{abstract}
Though type-Ia supernovae (SNe~Ia) are found in all types of galaxies, recent local Hubble constant measurements have disfavored using SNe~Ia in early-type or quiescent galaxies, aiming instead for better consistency with SNe~Ia in star-forming, late-type host galaxies calibrated by Cepheid distances. Here we investigate the feasibility of a parallel distance ladder using SNe~Ia exclusively in quiescent, massive ($\log M_*/M_\odot \geq 10$) host galaxies, calibrated by tip of the red giant branch (TRGB) distances. We present TRGB measurements to four galaxies: three measured from the \emph{Hubble Space Telescope} with the ACS F814W filter, and one measured from the \jwst NIRCam F090W filter. Combined with literature measurements, we define a TRGB calibrator sample of five high-mass, early-type galaxies that hosted well-measured SNe~Ia: NGC~1316 (SN~2006dd), NGC~1380 (SN~1992A), NGC~1404 (SN~2007on, SN~2011iv), NGC~4457 (SN~2020nvb), and NGC~4636 (SN~2020ue). We jointly standardize these calibrators with a fiducial sample of 124 Hubble-flow SNe~Ia from the Zwicky Transient Facility that are matched in host-galaxy and light-curve properties. Our results with this homogenized subsample show a Hubble residual scatter of under 0.11 mag, lower than usually observed in cosmological samples of the full SN~Ia distribution. We obtain a measurement of the Hubble constant, $H_0 = 75.3 \pm 2.9$ km s$^{-1}$ Mpc$^{-1}$, including statistical and estimated systematic uncertainties, and discuss the potential to further improve the precision of this approach. As calibrator and supernova samples grow, we advocate that future cosmological applications of SNe~Ia use subsamples matched in host-galaxy and supernova properties across redshift. \vspace{0.2in}
\end{abstract}

\keywords{\uat{Distance Indicators}{394} --- \uat{Galaxy Distances}{590} --- --- \uat{Hertzsprung Russell diagram}{725} --- \uat{Hubble constant}{758} --- \uat{Hubble Space Telescope}{761} --- \uat{Red giant tip}{1371} --- \uat{Standard Candles}{1563} --- \uat{Stellar Astronomy}{1583} --- \uat{Type Ia supernovae}{1728}}

\section{Introduction \label{sec:intro}}

At present, the most precise measurements of the current expansion rate of the Universe use a three-rung local distance ladder in which type-Ia supernovae (SNe~Ia) play a central role \citep[e.g.,][]{Riess2016,Riess2022,Dhawan2018,Freedman2019,Freedman2025}. The high and standardizable luminosities of SNe~Ia make them the best choice to extend the distance ladder into the smooth Hubble flow, where cosmological redshifts dominate observed recession velocities. Because SNe~Ia occur only once every few centuries in galaxies like the Milky Way, their luminosities cannot be calibrated directly through primary (geometric) distance indicators, requiring instead a middle rung to measure the Hubble constant. The best established of these secondary distance indicators are Cepheid variable stars \citep[e.g.,][]{Leavitt1912,Lee1993,Ferrarese2000,Riess2016,Riess2024} and the tip of the red-giant branch \citep[TRGB; e.g.,][]{Lee1993,Sakai1997,Bellazzini2001,Makarov2006,Rizzi2007,Jang2017b,Freedman2019,Freedman2021,Freedman2025,Scolnic:2023,Li:2025}. 

Normal SNe~Ia demonstrate broad homogeneity in their light curve properties \citep[e.g.,][]{Liu2023} with quantifiable variety that allows them to be standardized for use as precise cosmological distance indicators \citep[e.g.,][]{Phillips1993,Hamuy1996,Tripp1998,Guy2007,Jha2007,Riess2016,Kenworthy2021,Mandel2022}. SN~Ia light curves and luminosities (before and after standardization) are also correlated with host-galaxy properties, including host mass, star formation rate, and larger-scale environment \citep[e.g.,][]{Hamuy1995, Sullivan2010, Kelly2010, Sullivan2010, Burns2018, Rigault2020, Uddin2020, Larison2024, Ginolin:2025a, Ginolin:2025b, Senzel:2025}. 

The quest for increasing precision and accuracy in measuring \ho\ has made it imperative to limit systematic uncertainties in the distance ladder \citep{Freedman2021,Riess2022}. For SNe~Ia, this means ensuring that samples are consistent across the second and third rungs in all properties that could be correlated with luminosity. However, by necessity, the chosen secondary distance indicator determines the type(s) of SN~Ia host galaxies. Specifically classical Cepheids, being young stars, are only found in sufficient numbers in star-forming galaxies.

The TRGB method, conversely, can be applied to any system hosting stellar populations older than a few Gyr, requiring only a well-populated (and well-observed) red-giant branch (RGB). Specifically, it can be used to calibrate SNe~Ia in star-forming and quiescent host galaxies of all morphologies \citep[e.g.,][]{Mould2008,Tully2023}. However, the TRGB method has been limited by past observing facilities in its effective distance range ($D\leq20$~Mpc), restricting its application to more nearby galaxies that either can, or have already been, calibrated with Cepheids \citep[e.g.,][]{Riess2016,Beaton2019,Riess2024,Freedman2025}. Until recently, the Hubble Space Telescope (\hst) was the preferred observatory for TRGB-based distance measurements. The \jwst is now rapidly becoming the primary facility for the TRGB method based on its high angular resolution and increased sensitivity relative to the \hst. In particular, precise TRGB calibrations are already available for \jwst NIRCam wide filters \citep{Anand2024a,Newman2024a,Newman2024b,Hoyt2025,Li:2025}. Soon, the \jwst in combination with the upcoming Nancy Grace Roman Space Telescope (Roman) holds the potential to drastically increase the number of galaxies with TRGB-based distance measurements and ultimately improve the calibration of SN~Ia luminosities \citep[e.g.,][]{Anand2021, Kraemer2023}.

TRGB stars are evolved (4--12 Gyr), low-mass (0.8--2 $\msol$) stars that mark the end of the (first-ascent) red giant branch (RGB) stellar evolutionary phase and the onset of the helium flash. Due to the tight scaling relations between core mass, core radius, and core temperature of RGB stars, the bolometric luminosity at the TRGB is nearly uniform across stellar populations \citep{Serenelli2017}, but the luminosity at specific wavelengths can vary with stellar properties. Observationally, the TRGB appears as a sharp discontinuity and, given sufficient star counts, can be readily identified in a color-magnitude diagram \citep[CMD; e.g.,][]{DaCosta1990,Sakai1996,Madore1998}. In practice, the TRGB method is typically applied at I-band equivalent wavelengths where the brightness of the TRGB is only modestly dependent on mass, metallicity, and age \citep{Lee1993,Sakai1997,Beaton2019}. The \emph{HST} F814W filter, on both ACS or WFC3/UVIS, has been the standard TRGB filter choice for decades with continual improvements throughout, including updated zero-points, color-based metallicity corrections to the F814W luminosity function (LF), and more sophisticated methods to identify the TRGB in CMDs \citep[e.g.,][see \autoref{sec:trgb_method} for details;]{Ferrarese2000,Mendez2002,Bellazzini2001,Makarov2006,Rizzi2007,Mager2008,Freedman2021}. Since the launch of the \jwst, the TRGB method has been calibrated over a range of CMD combinations from the NIRCam F090W to the F444W filter. The TRGB feature in the \hst F814W and \jwst F090W filters exhibits a high degree of similarity in trends between the brightness versus metallicity and age \citep{McQuinn2019,Anand2024a,Newman2024b}. Thus, the \jwst F090W filter is currently preferred for calibrating rungs of the local distance ladder. There are significant gains in employing the TRGB method at NIR wavelengths where the TRGB stars appear up to 2 mags brighter relative to the F814W/F090W filters. These observations can extend the feasible distance range of the TRGB method by at least double and dramatically increase the volume, and thus the number of galaxies with secure TRGB distances \citep[see ][for NIR \jwst TRGB calibrations]{Newman2024b}. 

Because of the tension between the locally measured \ho\ and the inference derived from the cosmic microwave background (CMB) assuming standard $\Lambda$CDM, there is a compelling interest in checking each rung of the local distance ladder. Studies have compared TRGB and Cepheid distances to the same galaxies, particularly those that are SN~Ia hosts \citep[e.g.,][]{Freedman2021,Dhawan2023,AraucariaProject2023,Riess2024}. Such a comparison inherently restricts the sample to SN~Ia host galaxies in the intersection of both methods, specifically including only star-forming hosts and their subset of the overall SN~Ia population. This leaves out useful data, namely SNe~Ia found in non-star-forming host galaxies, calibrated through the TRGB. Here we focus on a parallel distance ladder using TRGB and SNe~Ia in massive, quiescent galaxies. 
 
Approximately 40\% of SNe~Ia in the low-redshift Universe are found in early-type host galaxies \citep[e.g.,][]{Li2011,Graur2017,Senzel:2025} and the majority of these host galaxies have stellar masses on the high side of the ``mass step'' seen in SN~Ia standardization, with $\log M_*/M_\odot \gtrsim 10$  \citep[e.g.,][]{Kelly2010,Uddin2020,Larison2024,Ginolin:2025a}. However, only two out of 22 SN~Ia host galaxies in the TRGB-calibrated second rung of the Carnegie-Chicago Hubble Program \citep[CCHP][]{Beaton2019,Freedman2019,Freedman2025} sample are massive, early-type hosts (NGC~1316 and NGC~1404). Separately, the TRGB-surface brightness fluctuation (SBF) program \citep{Anand2024a,Anand2025,Jensen2025}, designed to anchor the SBF distance technique to the \jwst TRGB and to measure \ho\, is incrementally increasing the number of massive, early-type galaxies through ongoing \jwst\ programs at distances beyond the reach of \emph{HST}. This includes the SN~Ia host NGC~1380 (SN~1992A). Recently, two SNe~Ia, SN~2020nvb and SN~2020ue, were observed in the massive, early-type galaxies NGC~4457 and NGC~4636, respectively. Both NGC~4457 and NGC~4636 are Virgo cluster galaxies located at a relatively large angular separation from the cluster center \citep[$8$\degree\ and $10.8$\degree, respectively,][]{Vollmer2013,Park2010}, and are morphologically classified as S0 and E/S0, respectively \citep{Vollmer2013,VeronCetty2006}. 

Here, we present a pilot study to measure TRGB distances to four nearby ($D\leq20$ Mpc), massive early-type SN~Ia host galaxies -- NGC~1316, NGC~1380, NGC~1404, and NGC~4457. We combine this with a TRGB distance for a similar SN~Ia host galaxy NGC~4636 \citep[taken from the literature as the data were still proprietary at the time we unblinded our analysis]{Anand2025}. We then apply this TRGB calibration to a sample of Hubble-flow SNe~Ia with matched supernova light-curve and host-galaxy properties from the ZTF SN~Ia Data Release 2 \citep{Rigault:2025a} producing a ``proof-of-concept'' measurement of the Hubble constant.

In \autoref{sec:observations}, we present new \hst observations for NGC~4457, archival \hst data for NGC~1404 and NGC~1316, and archival \jwst data for NGC~1380,introduce our data reduction methods, including image alignment, photometry, artificial star tests, and high-fidelity photometric catalog culling, and describe our SN~Ia observations and data reduction process. In \autoref{sec:distance_methods}, we detail and apply our methodology for TRGB-based distances, while the SN~Ia light curve fitting and standardization methodology is described in \autoref{sec:snIa}. In \autoref{sec:H0}, we discuss our fiducial calibrator and Hubble-flow SN~Ia sample selection, describe our joint standardization and cosmological model fitting routine, and present the fiducial \ho, along with results from several sample selection variants. Finally, in \autoref{sec:conclusion}, we discuss and summarize our findings, including our recommendations for increasing the sample with \emph{JWST}.

\section{TRGB Observations and Data Reduction}\label{sec:observations}
Our galaxy sample includes the few early-type galaxies that host at least one SN~Ia with well-sampled light curves and have the requisite observations for a precise TRGB measurement: NGC~1316, NGC~1380, NGC~1404, NGC~4457, and NGC~4636.

\subsection{Reduction of Imaging Data}
\begin{figure*}[!htb]
    \centering
    \includegraphics[width=0.45\textwidth]{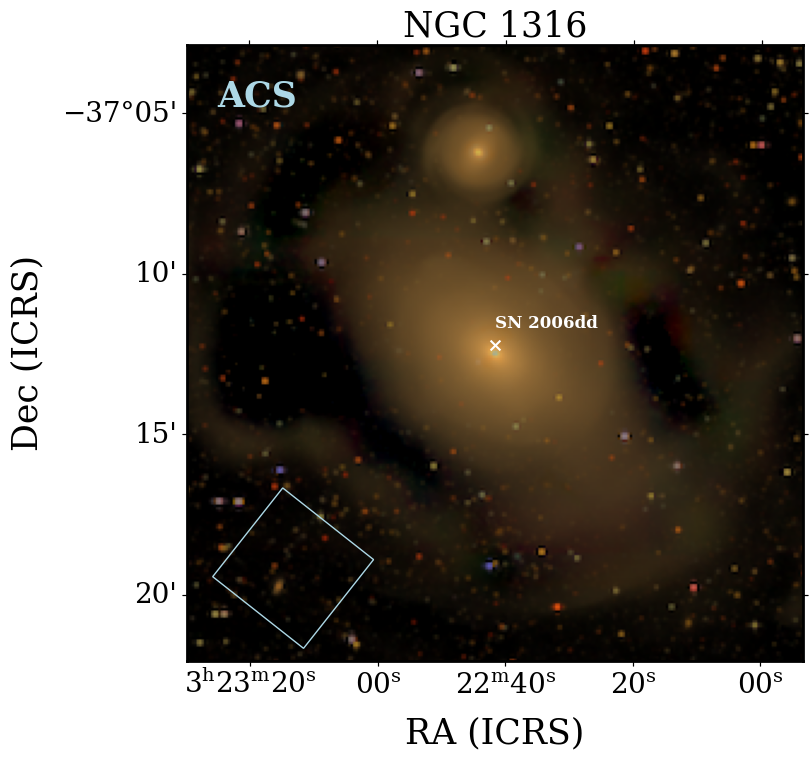}
    \includegraphics[width=0.45\textwidth]{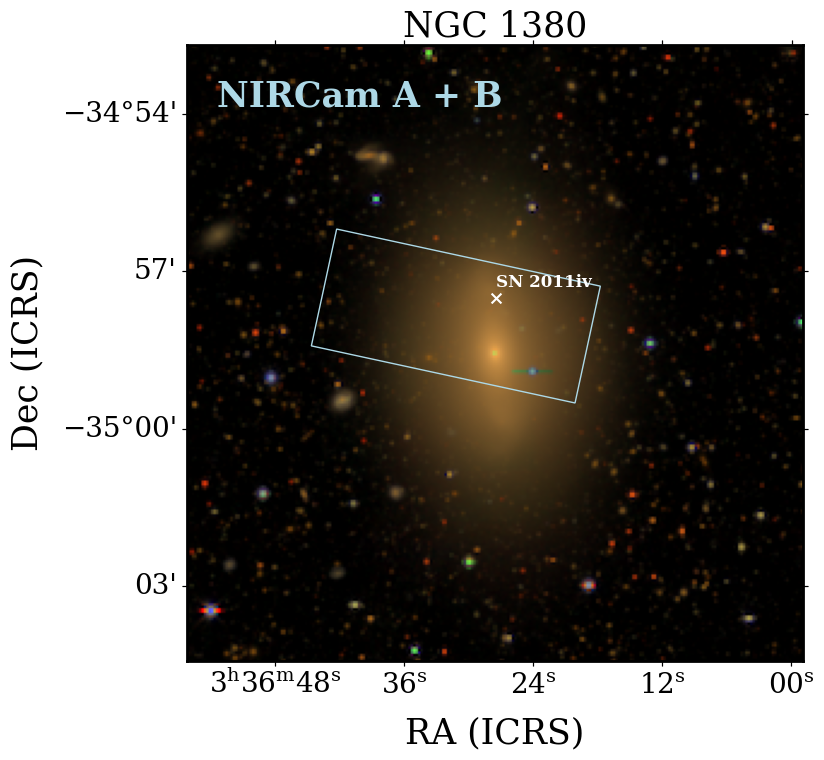}
    \includegraphics[width=0.475\textwidth]{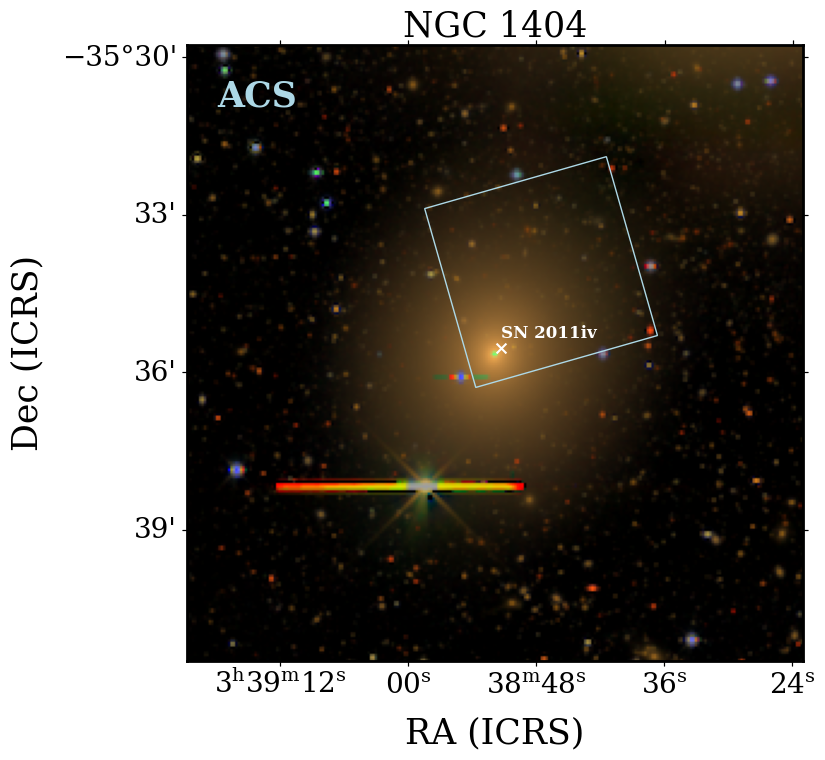}
    \includegraphics[width=0.45\textwidth]{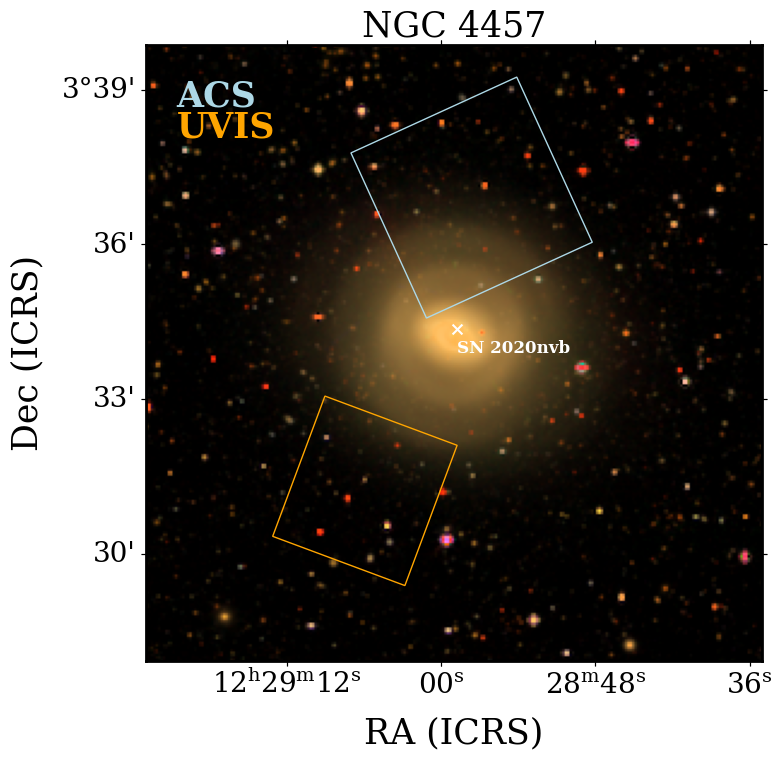}
        \caption{The observed fields for TRGB distance measurements for each galaxy in this study. Clockwise from the top left are NGC~1316, NGC~1380, NGC~1404, and NGC~4457. Background images are 3-color Legacy Survey Data Release 10 \citep[LS-DR10;][]{Dey2019} images where we assign the $g$, $r$, and $i$ bandpasses to the green, blue, and red channels, respectively. Overlaid on images are either the ACS (NGC~1316, NGC~1404, and NGC~4457) or the NIRCam Instrument (NGC~1380) footprints (light blue) and the location and name of the SN~Ia (black cross). The UVIS (orange) footprint is additionally overlayed on the NGC~4457 image. Images are all oriented with north up and east left. The LS-DR10 images were retrieved via url from the legacy survey web host \citep{Dey2019}.}
    \label{fig:sample_images}
\end{figure*}

New observations of NGC~4457 were acquired with the \hst Advanced Camera for Surveys Wide Field Camera (ACS/WFC; hereafter ACS) as part of \hst-GO-16453 (PI McQuinn). We optimized the observations for TRGB measurements by imaging the outer stellar fields of the galaxy with the F814W and F606W filters. Coordinated parallel observations were obtained with the Wide Field Camera 3 imager ultraviolet and visible light channel (WFC3/UVIS; hereafter UVIS) F606W and F814W filters. Shown in the right panel of \autoref{fig:sample_images} are the locations for our ACS (light blue) and UVIS (orange) observations.

Both NGC~1316 and NGC~1404 have deep archival ACS observations in F606W and F814W obtained for the CCHP \citep[see][respectively]{Hatt2018,Hoyt2021}. The left and middle panels of \autoref{fig:sample_images} show the ACS pointings for NGC~1316 and NGC~1404, respectively. We obtained the ACS exposures for NGC~1316 and NGC~1404 from the Mikulski Archive for Space Telescopes (MAST;\dataset[doi:10.17909/T9RP4V]{https://dx.doi.org/10.17909/T9RP4V}). 

NGC~1380 and NGC~4636 were both observed with the \jwst in the Near Infrared Camera (NIRCam) detector imaging mode in the program JWST-GO-3055 \citep[The TRGB-SBF Project;][]{Anand2024a, Anand2025}. For optimal TRGB distance measurements, the observing strategy employed the F090W filter and imaged galaxy fields beginning outside the dense galactic stellar discs and extending out to the galactic halos. At the time of this study, only NGC~1380 has publicly available archival data on MAST; NGC~4636 observations were still in their exclusive access period at the time of unblinding. We therefore check the consistency between the distance scale presented in \citet{Anand2024a} and this study by independently reducing NGC~1380 observations and measuring a TRGB distance with our own methods. We then adopt the TRGB distance for NGC~4636 presented in \citet[][see \autoref{sec:trgb_method} for details]{Anand2025}. In \autoref{tab:observations_summary}, we provide several summary statistics for observations of the five galaxies.

\begin{deluxetable*}{|l|cccccc|}
\tablewidth{0pt}
    \tablecaption{Summary of the \hst and \jwst Galaxy Sample Observations \label{tab:observations_summary}}    
    \tablehead{
    \colhead{Galaxy} & \colhead{R.A. (J2000)} & \colhead{Decl. (J2000)} & \multicolumn{2}{c}{Exposure Time (s)} & \colhead{PID} & \colhead{Milky Way $E(B-V)$}\tablenotemark{a}\\
     \nocolhead{}       &  \colhead{(h:m:s)}    &  \colhead{(d:m:s)} & \multicolumn{2}{c}{Filters} & \nocolhead{} & \colhead{(mag)}
    }
    \startdata
    \multicolumn{3}{|c}{} & F606W & F814W & \multicolumn{2}{c|}{}\\
    \hline
    NGC~1316 & 03:23:12.69 &  $-$37:19:19.33& 14400 & 24000 & HST-GO-13691 & 0.021\\
    NGC~1404 & 03:38:46.68& $-$35:34:04.08 & 40800& 40800 & HST-GO-15642 & 0.012 \\
    NGC~4457 & 12:28:57.59 &  $+$03:36:56.24 & 8200 & 8200 & HST-GO-16438 & 0.022\\
    \hline
    \hline
    \multicolumn{3}{|c}{} & F090W & F150W & \multicolumn{2}{c|}{}\\
    \hline 
    NGC~1380 & 03:36:27.59& $-$34:58:34.68& 7730 & 1933 & JWST-GO-3055 & 0.017 \\
    NGC~4636 & 12:42:49.833 & $+$02:41:15.95 & 5154 & 1074 & JWST-GO-3055 & 0.029
    \enddata
    \tablenotetext{a}{Reported $E(B-V)$ values are sampled from the \citet{Schlafly2011} recalibration of the \citet{Schlegel1998} dust map at R.A. and Decl. coordinates listed in columns 1 and 2.}
\end{deluxetable*}

\subsubsection{Image Alignment}\label{sec:image_alignment}
 Precise alignment of \hst observations is essential to measuring robust photometric properties of sources in the images. We use the tool \TweakReg{} included in the \Drizzlepac{} Python package to perform image alignment and correct world coordinate system (WCS) solutions for individual exposures. First, we generate separate, matched source catalogs for each filter. Second, we run \TweakReg{} on the combined source catalog to update all WCS solutions to a single reference frame with the constraint that, at minimum, 10 sources are cross-matched in all images. WCS solutions are considered satisfactory when the alignment's root mean square (RMS) is better than 0.1 pixels.
 
 Reference images for the ACS (or UVIS where applicable) F606W and F814W filters were generated using the \Astrodrizzle{} software included in \Drizzlepac{}. Here, all exposures in a given filter are combined into a single, deep image (i.e., a drizzled image). We use the F606W image as our final reference frame in the photometry processing (\autoref{sec:phot}). 

 Alignment solutions for the \jwst NIRCam observations of NGC~1380 were taken directly from the pipeline as they were sufficient for this work. We selected as a reference image the mosaiced \texttt{i2d extension} image in the F090W filter. 

\subsubsection{Photometry} \label{sec:phot}
 We performed point-spread function (PSF) photometry on the well-aligned \hst images to generate the data for our analysis. We use the software \Dolphot{}, a modified version of the WFPC2-specific package \hstphot which provides ACS, UVIS, and \jwst NIRCam-specific modules \citep{Dolphin2002,Dolphin2016,Weisz2024}. As mentioned above for \hst (see \autoref{sec:image_alignment}), we set the drizzled F606W image as our reference frame for source identification owing to its high angular resolution. \Dolphot{} also requires a parameter file with global and image-specific values. We adopted the values from the Panchromatic Hubble Andromeda Treasury (PHAT) and Panchromatic Hubble Andromeda: Triangulum Extended Region (PHATTER) surveys \citep[see][]{Williams2014,Williams2021}. We use the parameter values for the short wavelength photometry from the \jwst Early Release Science Resolved Stellar Populations Program \citep{Weisz2024}.

\subsubsection{Artificial Star Tests}
We use artificial star tests (ASTs) to quantify photometric completeness (recovery fraction) and photometric errors due to blending in our stellar catalogs. We injected $\sim100,000$ artificial stars into each image. The spatial distribution of these artificial stars spans the full range of the sources determined in the photometry (\autoref{sec:phot}). Fake star coverage of the CMDs is uniformly distributed to span the entire parameter space where real sources appear. We then run photometry on all fake stars in each image using \Dolphot{} with identical parameter files to the original runs. The results of the ASTs are used directly in our TRGB fitting method (see \autoref{sec:trgb_method}).

\subsubsection{High-fidelity Photometric Catalogs}\label{sec:SpatialCut}
To optimize our stellar catalogs for robust TRGB measurements, we apply two different methods to cull the initial photometry: thresholds on how well an individual source is recovered and spatial cuts that follow each galaxy's structural parameters. We find this combination highly effective in generating high-fidelity stellar catalogs for our galaxies.

We start by applying conservative, per filter cuts to the full photometric output. \Dolphot{} provides important quality metrics for \emph{every} source it identifies in the image. In particular, we focus on the crowding metric, a measure in magnitudes of how much brighter a star would be in the absence of nearby sources; the sharpness metric, a measure of how point-like or extended an object is (negative values are typically cosmic rays, a value of zero is a perfectly fit source, and positive values correspond to extended sources); and the signal-to-noise ratios (SNRs) \citep{Dolphin2002, Dolphin2016, Weisz2024}. 

We first retain from the catalog only sources with lenient quality metric thresholds: crowding values of $\leq1$ mag; sharpness between $-0.316$ and $0.316$ (sharpness$^2\leq0.1$); and SNRs of $\geq2$ (non-TRGB magnitude filter, F606W or F150W) and F814W or F090W $\geq4$. We then examine the remaining sources, iteratively tightening the crowding and sharpness$^2$ cuts until no visual diffraction spike artifacts remain in the catalogs. These initial cuts are used to remove non-stellar sources and/or poorly recovered sources. 

Next, we consider how the spatial distribution of sources in our targets impacts the quality of their photometry. The targeted fields in our galaxy sample's \hst observations are not uniformly placed relative to the galaxy's center. We targeted two different fields around NGC~4457, with the ACS field as the primary. Part of the ACS field covers the outer stellar field, while the other part is nearer to the dense central nucleus. The UVIS field is almost entirely positioned in the outer stellar field. The \hst observations for NGC~1316 are firmly in the outer stellar disk. The \hst observations for NGC~1404 are placed similarly to the ACS field in NGC~4457; however, the NGC~1404 field overlaps directly with the galaxy's center. For NGC~1380 the \jwst\ NIRCam instrument pointing covers the central region and an outer stellar disk region in NIRCam A and NIRCam B, respectively. The observing design is identical for NGC~4636.

\begin{figure*}
\centering
    \includegraphics[width=0.47\linewidth]{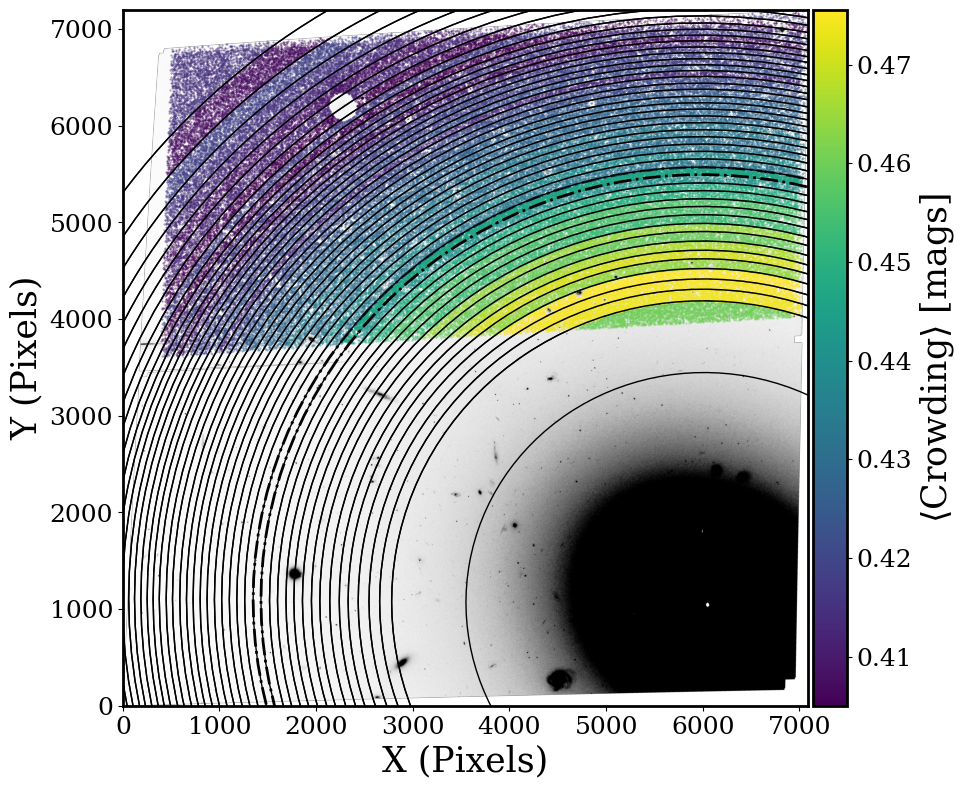}
    \includegraphics[width=0.39\linewidth]{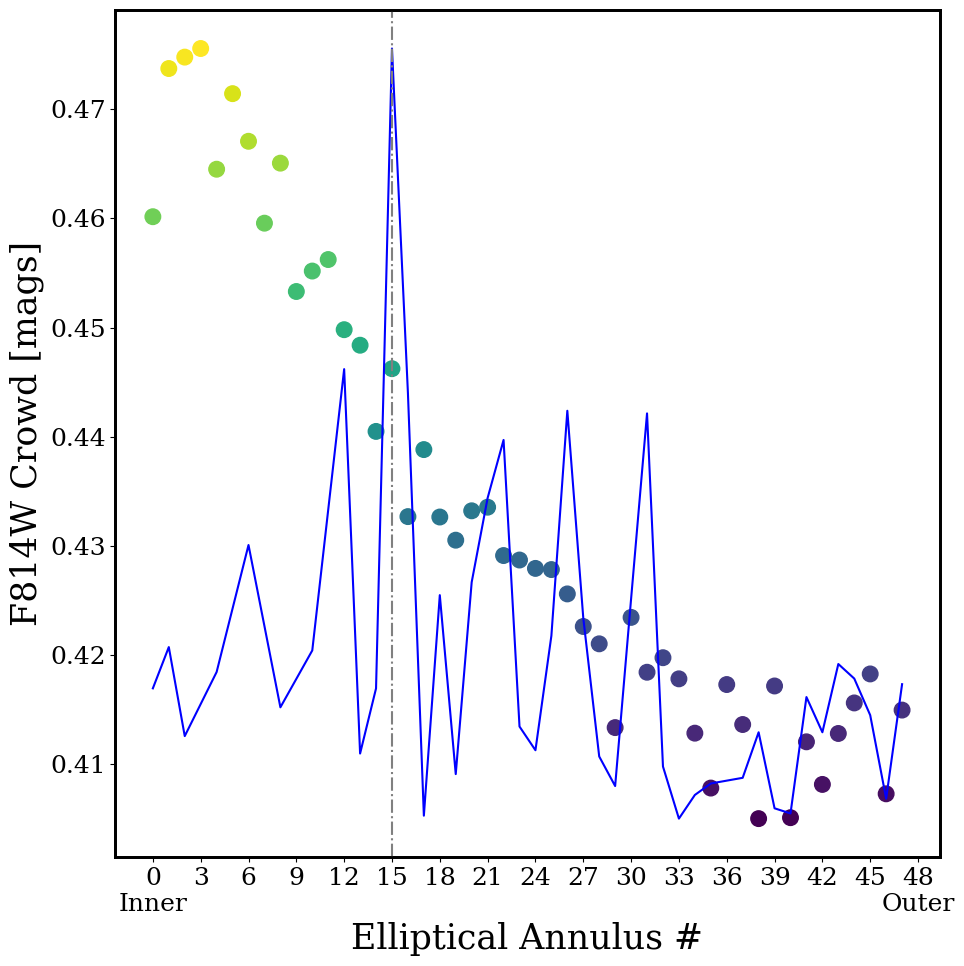}
    \caption{We demonstrate our high-fidelity photometry catalog culling method applied to NGC~1404 to identify the appropriate spatial region and crowding parameter threshold. We include only photometry for NGC~1404 from ACS chip 2 as a result of chip 1 producing suboptimal photometry due to the highly crowded center of NGC~1404. Left: Individual points are stars in our initial photometry catalog. The cut-off for each concentric elliptical annuli (solid black curves) is determined iteratively with the constraint of until each annulus has approximately equal numbers of sources. Points are color-coded by the mean crowding value (see the color bar for values) within an annulus. The cut-off ellipse is indicated as a dash-dotted line. Right: Mean crowding value versus annulus number. The points indicate mean crowding per annulus where the color-coding is identical to the left panel. We apply a discrete first-derivative filter to the individual points (blue curve) and identify the largest peak as the innermost location from the center of the galaxy that constitutes our high-fidelity stellar catalog for TRGB measurements.}
    \label{fig:spatial_selection}
\end{figure*}

For each galaxy, we follow the same procedure. First, we divide our initial source catalog into concentric elliptical annuli about the center of a galaxy. We adopt the structural parameters, namely ellipticity, position angle, and semi-major and minor axes, from the literature. These parameters and their references are provided in \autoref{tab:structure_params} for each galaxy. Concentric annuli are generated iteratively with approximately equal numbers of sources in each annulus. In \autoref{fig:spatial_selection}, left panel, we present an example of our culling method applied to the ACS field of NGC~1404. Concentric annuli (solid black curves) are over-plotted on the x/y locations of the source catalog in pixel coordinates. Second, we calculate the mean crowding value within each annulus. The sources are color-coded by their mean crowding value (see the color bar). Third, we apply a discrete derivative filter to the mean crowding values as a function of the annulus number. In \autoref{fig:spatial_selection}, right panel, we show the mean crowding versus annulus number (identical color-coding as in the left panel) and the derivative response (blue curve). We identify the peak in the derivative as the location of our spatial and crowding value cut-off.  
 
\begin{deluxetable}{l|cccD}
    \tablewidth{0pt}
    \tablecaption{Summary of Galaxy Structural Parameters}
    \label{tab:structure_params}
    \tablehead{
    \colhead{Galaxy} & \colhead{R.A. (J2000)} & \colhead{Decl. (J2000)} & \colhead{P.A.} & \colhead{$e$} \\
     \nocolhead{empy}   &  \colhead{(deg)}  &  \colhead{(deg)} &  \colhead{(deg)} & \nocolhead{$\ldots$}
           }
    \startdata
    NGC~1316 & 50.67  & $-$37.08 & $+$55.4 & 0.3\\
    NGC~1380 & 54.11  & $-$34.98 & $-$84.1 & 0.6 \\
    NGC~1404 & 54.72  & $+$35.59 & $-$10.3 & 0.2\\
    NGC~4457 & 187.25 & $+$03.57 & $+$70.5 & 0.8
    \enddata
    \tablecomments{PA (Position Angle); $e$ (ellipticity) $=1-\left(b/a\right)$. We adopt structural parameters from the literature for NGC~1316 \citep[DES DR1;][]{Abbott2018}, NGC~1380 and NGC~1404 \citep[DES DR2;][]{Abbott2021}, and NGC~4457 \citep[SDSS DR9;][]{Adelman-McCarthy2012}. R.A.\ and Decl.\ indicate the galaxy center on which elliptical annuli are centered, and are different than the coordinates of the telescope pointings listed in \autoref{tab:observations_summary}.}
\end{deluxetable}

\begin{figure*}[!h]
\centering
    \includegraphics[width=0.335\textwidth, trim=0 0.4cm 0 0]{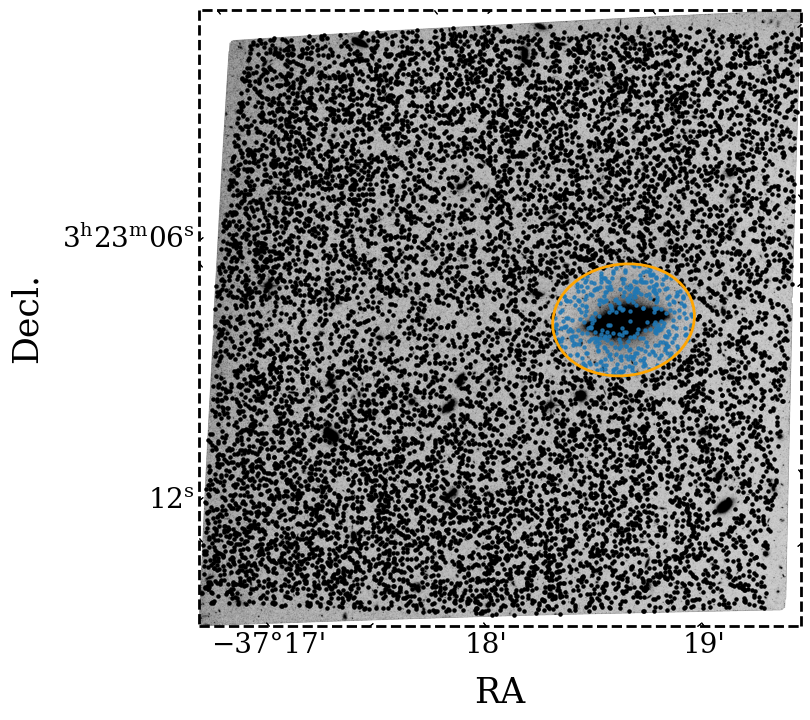}
    \includegraphics[width=0.3\textwidth]{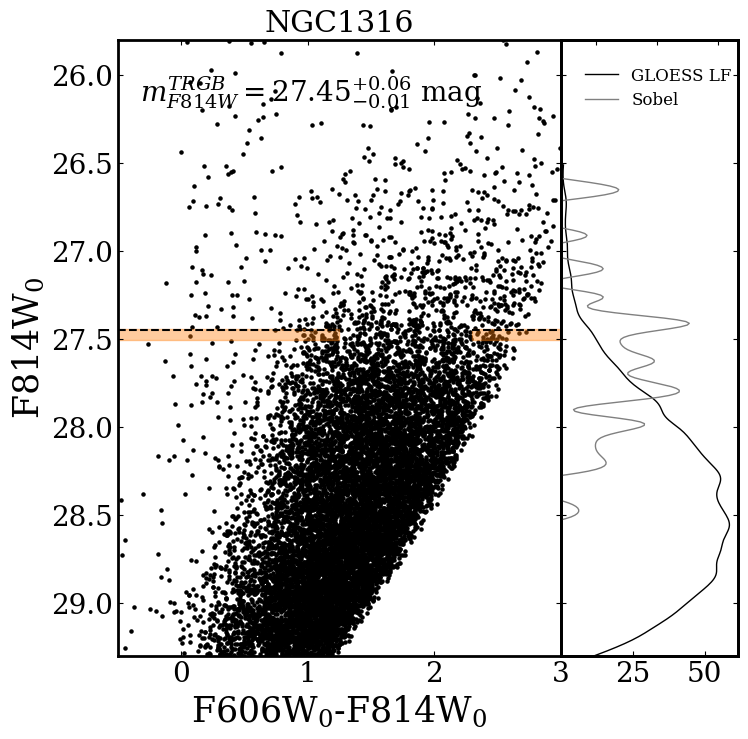}
    \includegraphics[width=0.3\textwidth]{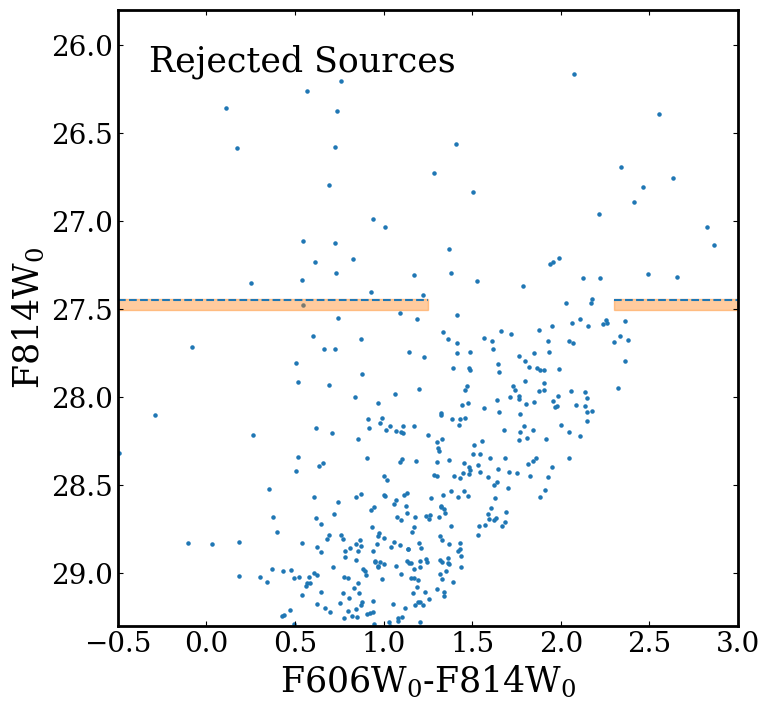}
    \caption{The high-fidelity extinction-corrected stellar catalog for NGC~1316. Left: The spatial distribution of stellar sources included (excluded; see \autoref{sec:SpatialCut}) in the TRGB fit is shown as black (light blue) points. Center: The final high-fidelity CMD from which we measure the TRGB magnitude. The horizontal black line marks the location of the TRGB with the uncertainty band (orange-shaded region). Right: The CMD of stars within the spatial region that we excluded from our analysis. We indicate the TRGB location for reference. We report a TRGB magnitude of $m^{TRGB}_{F814W}=27.45^{+0.06}_{-0.01}$ mag.}
    \label{fig:ngc1316_trgb}
\end{figure*}

\begin{figure*}[!h]
\centering
    \includegraphics[width=0.355\textwidth, trim=0 5 0 2]{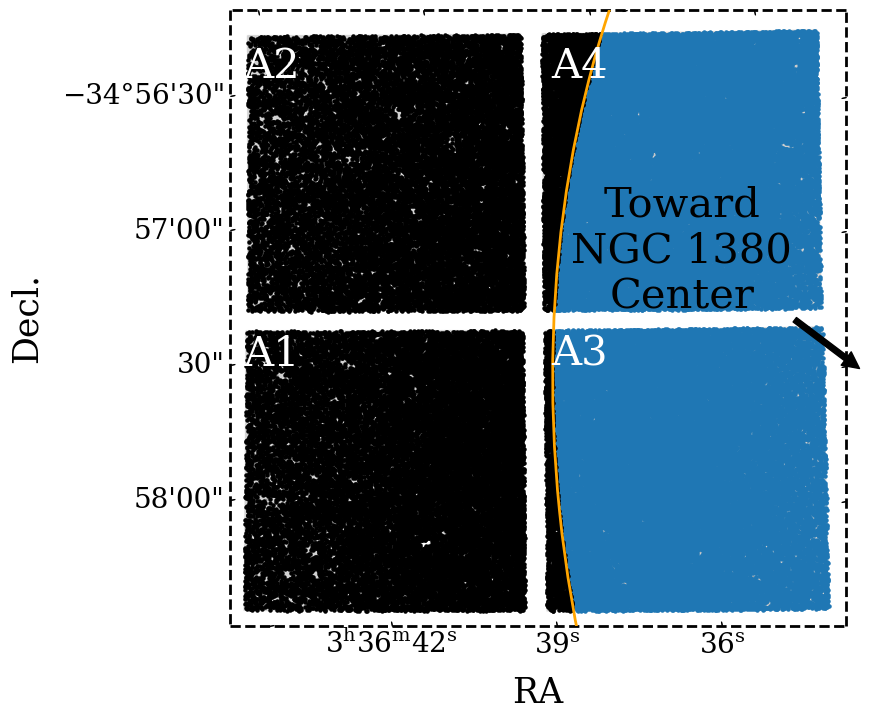}
    \includegraphics[width=0.3\textwidth]{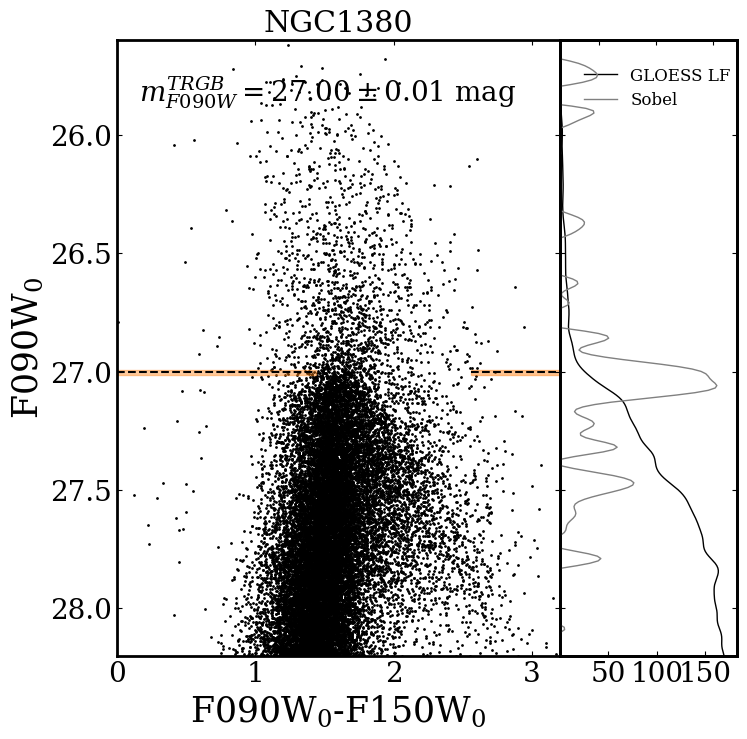}
    \includegraphics[width=0.3\textwidth]{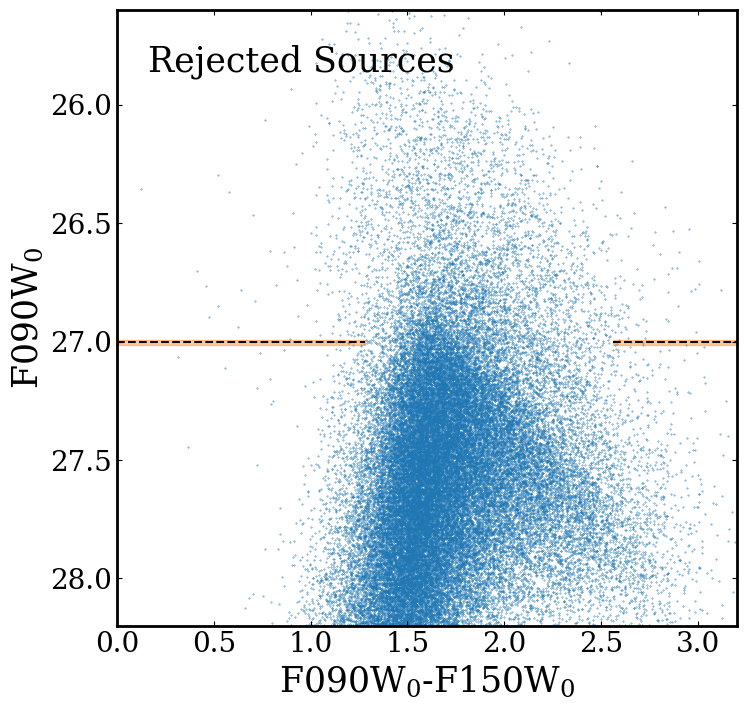}
    \caption{Similar to \autoref{fig:ngc1316_trgb} but now for NGC~1380. The left panel shows the NIRCam A (NRCA) with the four Short wavelength (SW) detectors A1, A2, A3, and A4. The inner detectors, A3 and A4, are positioned closest NGC~1380's center. The spatial cuts removed a majority of stars in the inner two detectors. Our final TRGB measurement for NGC~1380 is $m^{TRGB}_{F814W}=27.00\pm0.01$ mag. This is in close agreement with the distance modulus reported in \citet{Anand2024a}. Note that \citet{Anand2024b} reported photometry and subsequent analysis in the previous Vega-Vega photometric flux calibration as opposed to this study for which we use the current flux calibration tied to the star Sirius. We therefore compare in distance moduli rather than the TRGB magnitude directly since differences in magnitudes are irrelevant to the distance moduli (see \autoref{sec:trgb_method} for additional details).}
    \label{fig:ngc1380_trgb}
\end{figure*}

\begin{figure*}
\centering
    \includegraphics[width=0.34\textwidth, trim=0 0.4cm 0 0]{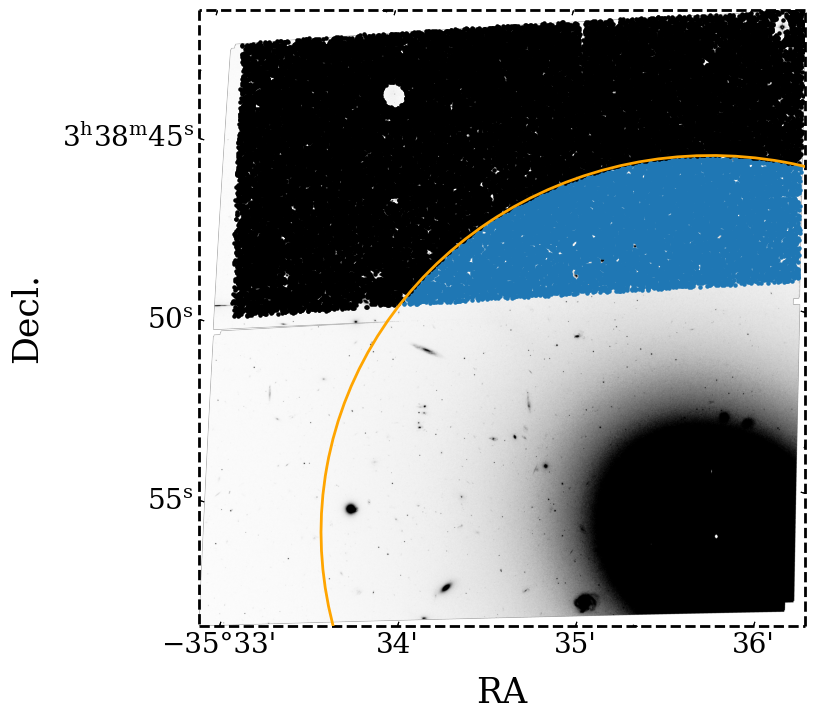}
    \includegraphics[width=0.32\textwidth]{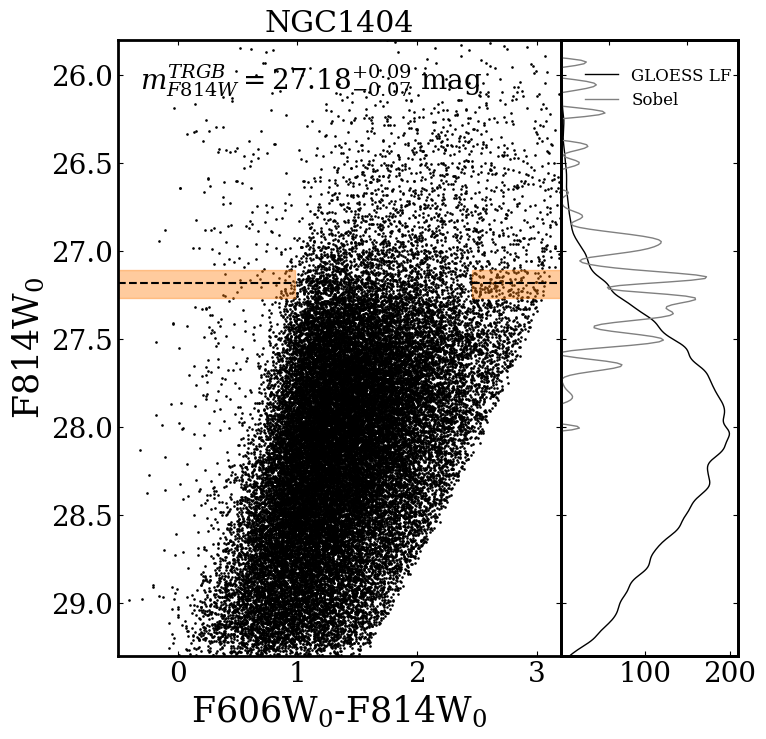}
    \includegraphics[width=0.32\textwidth]{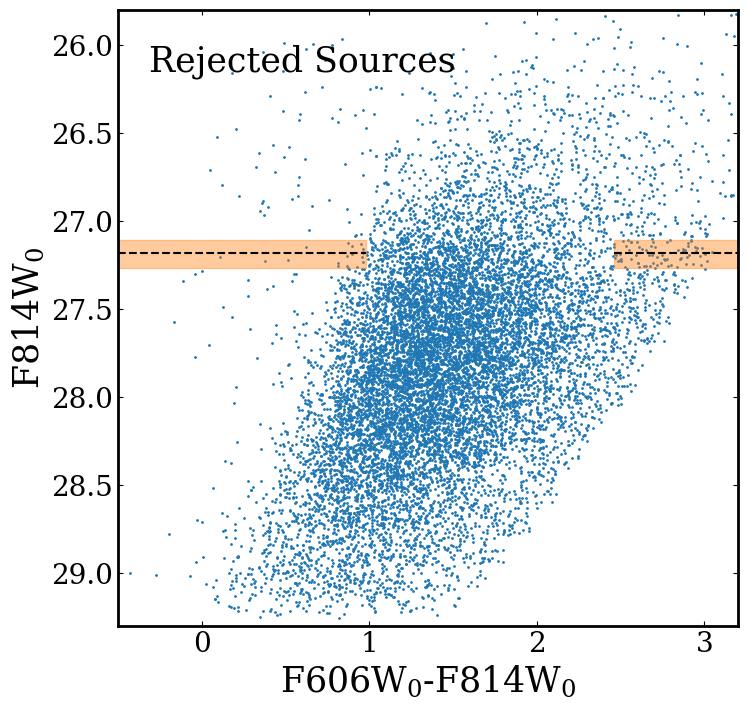}
    \caption{Similar to \autoref{fig:ngc1316_trgb} but now for NGC~1404. We excluded from our high-fidelity stellar catalog nearly the entire ACS/WFC Chip 2 due to increased saturation in NGC~1404's bright and dense nucleus. We also excluded sources in the region around the dwarf galaxy FCC B1281 that is at the same approximate distance as NGC~1404 \citep{Hoyt2021}. We report a TRGB magnitude of $m^{TRGB}_{F814W}=27.18^{+0.09}_{-0.07}$ mag.}
    \label{fig:ngc1404_trgb}
\end{figure*}

\begin{figure*}[!tbh]
    \centering
    \includegraphics[width=0.34\textwidth, trim=0 0.3cm 0 0]{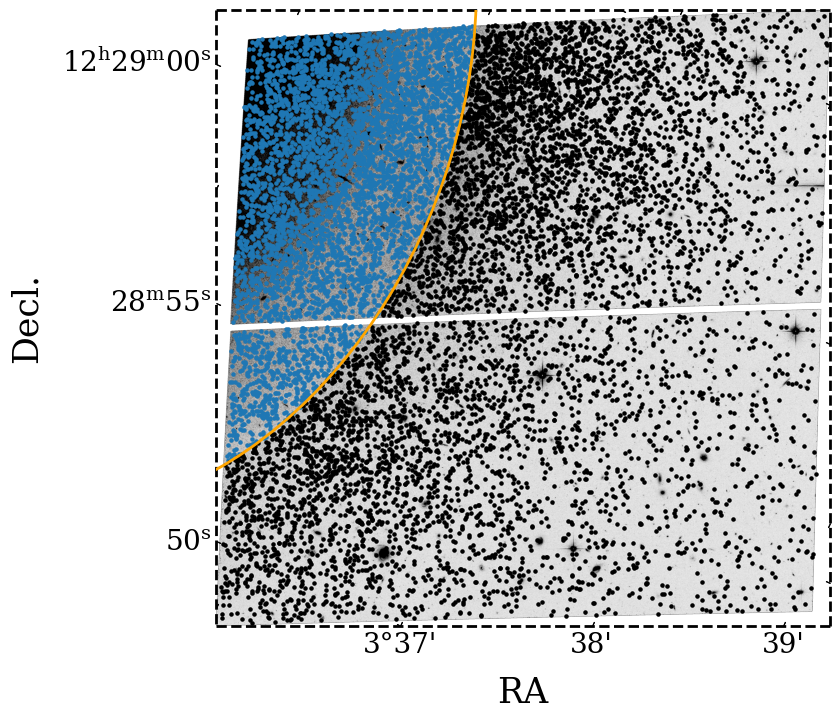}
    \includegraphics[width=0.3\textwidth]{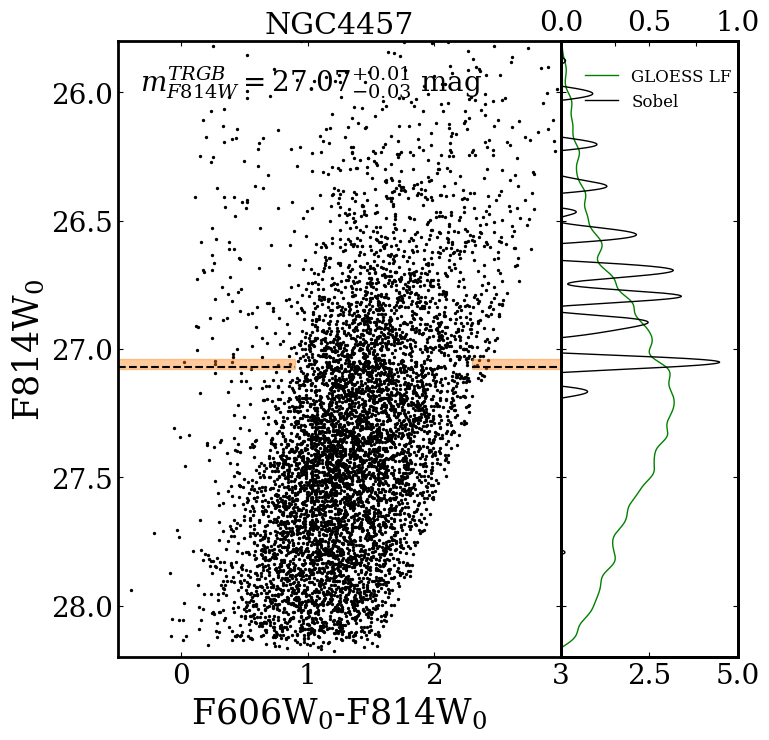}
    \includegraphics[width=0.31\textwidth]{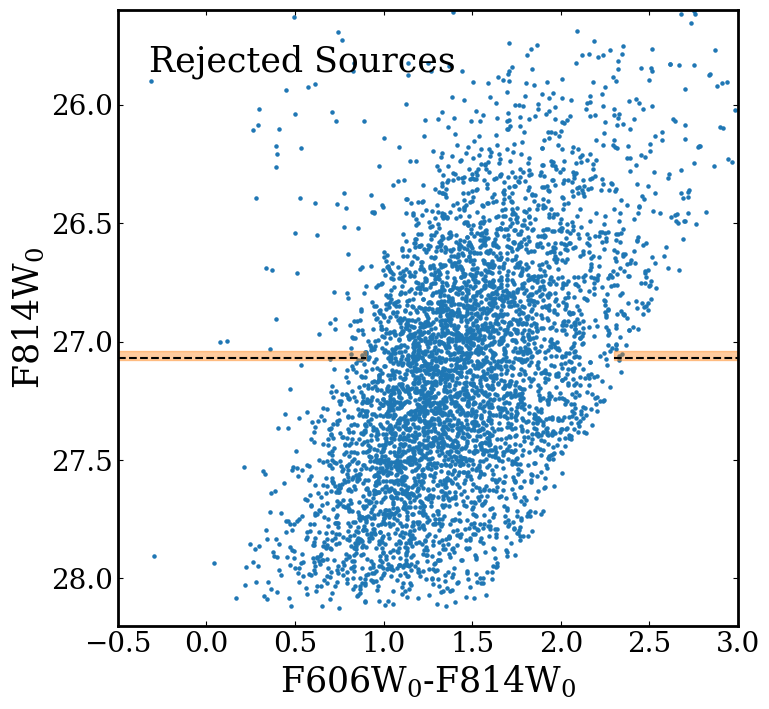}
    \caption{Similar to \autoref{fig:ngc1316_trgb} but now for NGC~4457. We present ACS-only photometry (not UVIS; see Appendix~\autoref{sec:ngc4457_uvis_dist}. We report a TRGB magnitude of $m^{TRGB}_{F814W}=27.07^{+0.01}_{-0.03}$ mag.}
    \label{fig:ngc4457_acs_trgb}
\end{figure*}

Finally, in the left panels of \autoref{fig:ngc1316_trgb} to \autoref{fig:ngc4457_acs_trgb}, we show the spatial distribution of high-fidelity sources (black points) and sources excluded from further consideration (blue points). The final \Dolphot{} quality metric thresholds are summarized in \autoref{tab:hifid_cuts}. These thresholds include the number of stars, first with only the strict quality metric cuts applied ($N_*$) and second, including both the quality metric and spatial cuts ($N_{*,\text{SMA}\geq \text{SMA}_{\text{inner}}}$).  We note two additional spatial considerations we made in creating our high-fidelity catalogs. First, as shown in the left panel of \autoref{fig:ngc1316_trgb}, we excluded from the NGC~1316 catalog sources in the region around the background galaxy 2MASS-J03231562-3719444. Second, shown in the left panel of \autoref{fig:ngc1404_trgb}, we excluded from the NGC~1404 catalog sources about the nearby background dwarf galaxy FCC B1281. Sources excluded from the catalog are also shown in CMD space in \autoref{fig:ngc1316_trgb} through \autoref{fig:ngc4457_acs_trgb}.

\begin{deluxetable}{|c|cccc|cc|}
    \tablewidth{0pt}
    \tablecaption{Summary of High-fidelity Photometric Source Catalog Properties}
    \label{tab:hifid_cuts}
    \tablehead{\colhead{Galaxy} &
    \colhead{Crowding (mag)} & \colhead{Sharpness$^2$} & \colhead{SNR (Blue, Red)} &
    \colhead{$N_*$} & \colhead{SMA$_{\text{inner}}$} & \colhead{$N_{*,\text{SMA}\geq \text{SMA}_{\text{inner}}}$}
    }
    \startdata
    \hline
    NGC~1316 & $0.15$ & $0.03$ & $2.0$, $4.0$ & 12,035 & \nodata & \nodata  \\
    NGC~1380 & $0.15$ & $0.02$ & $4.0,4.0$        & 110,556 & $201.81''$ & 47,743 \\
    NGC~1404 & $0.43$ & $0.03$ & $2.0$, $4.0$ & 52,091 & $138.69''$ & 39,486 \\
    NGC~4457 & $0.19$ & $0.02$ & $2.0$, $4.0$ & 11,445 & $156.51''$ & 6,661 
    \enddata
    \tablecomments{The quality metric thresholds, full field star counts, inner semi-major-axis (SMA, arseconds) beyond which stars are retained, and star counts after spatial clipping are reported for each target. Crowding and sharpness$^2$ are upper limits while the SNR per filter are lowers limits. For all \hst -observed targets we find two different SNR thresholds, one more lenient in F606W, provides a satisfactory color baseline in the CMDs for measuring the TRGB. The total number of high-fidelity sources are determined after applying quality cuts and removing sources that fall in the excluded spatial region (where applicable).}
\end{deluxetable}

\section{TRGB Distance Measurements: Methods \& Results}\label{sec:distance_methods}
This section describes the TRGB distance measurement technique and summarizes our findings.

\subsection{TRGB Methodology}\label{sec:trgb_method}
Since its first modern application by \cite{Lee1993}, the techniques used to measure the location of the TRGB feature in CMDs have increased in quantity and sophistication. Early in its history, the apparent magnitude of the TRGB was often identified visually in a CMD. \citet{Lee1993} introduced a formalized method for identifying the TRGB by first generating an I band LF (i.e., a 1-D histogram) marginalized over the color of the stars in a CMD and second convolving an edge-detection filter, a zero-sum Sobel kernel of the form $[-2, 0, +2]$. The Sobel kernel has since been revised in several ways to account for sources of uncertainty that can act to shift the location of the TRGB in a LF \citep[e.g., various forms of Gaussian smoothing,][]{Sakai1996,Hatt2018,Beaton2019}. 

As an alternative to the Sobel-kernel approach, a more sophisticated Bayesian maximum likelihood -based technique was developed to measure the TRGB. This technique includes a parametric form for the RGB LF, which it uses to fit the observed LF \citep[see][]{Mendez2002, Makarov2006}. The probability distribution includes terms that account for the photometric uncertainty distribution and completeness function from ASTs \citep[see][]{Makarov2006}. For our TRGB measurements, we adopt the same theoretical LF form used in \citet{Makarov2006}:
\begin{equation}
P = 
\begin{cases}
	10^{\left(A*\left(m-m_{\text{TRGB}}\right)+B\right)},  & \text{if } m-m_{\text{TRGB}} \geq 0 \\
	10^{\left(C*(m-m_{\text{TRGB}})\right)},  & \text{if } m-m_{\text{TRGB}} < 0 
\end{cases}
\label{eq:ml_form}
\end{equation}
where A characterizes the RGB slope with a normal prior of $\mu_{\text{A}}=0.3$ and uncertainty of $\sigma_{\text{A}}=0.07$, B represents the RGB jump, and C is the AGB slope with a normal prior of $\mu_{\text{C}}=0.3$ and an uncertainty of $\sigma_{\text{C}}=0.2$. Parameters A, B, and C are free parameters in our application. The maximum likelihood measurement uncertainty is based on the range of solutions returning the log of the probability, $\log{(P)}$, within 0.5 of the maximum. 

In addition, we consider and account for the extended stellar metallicity/age in our galaxy sample. It is well-established that the TRGB brightness is approximately constant in F814W for only the oldest and most metal-poor RGB stars. At higher metallicities and/or younger ages, the TRGB deviates from a constant brightness \citep[e.g.,][]{Lee1993,Rizzi2007}. In practice, this metallicity/age effect is traced by the color of the stars. We correct for this color-based metallicity dependence and adopt the calibration from \citet{Jang2017b}, the quadratic transformation (QT) system, derived for the \hst filters \citep[see also][]{Hoyt2021}. We reproduce the calibration here, over the applicable color range F606W$-$F814W$>1.5$, for convenience:

\resizebox{1.0\linewidth}{!}{
	\begin{minipage}{1.2\linewidth}
        \begin{align}\label{eqn:Jang2017b}
            \text{F814W}_{\text{corrected}} = \text{F814W}_{\text{fiducial}} 
            &-a_{\text{quadratic}}\left[\left(\textrm{F606W-F814W}\right)-1.1\right]^2 \\\nonumber
            &+b_{\text{linear}}\left[\left(\textrm{F606W-F814W}\right)-1.1\right] \\ \nonumber
        \end{align}
\end{minipage}
}
where F814W$_{\text{fiducial}}$ is the original apparent magnitude of the stars, F814W$_{\text{corrected}}$ is the color-corrected magnitude (i.e., the rectified magnitude), and $a_{\text{quadratic}} = 0.159\pm0.025$ and $b_{\text{linear}}=-0.047\pm0.039$ are the best-fit parameters for the quadratic and linear terms, respectively. The TRGB is found to be constant in brightness for colors F606W-F814W$\leq1.5$ mag; therefore, we apply no correction for stars with colors blueward of 1.5 mag. The stars with colors F606W-F814W$>1.5$ mag are color-corrected to the pivot point of F606W-F814W$=1.1$ mag where the TRGB is still flat, and thus, the apparent magnitude is constant. \citet{Jang2017b} note that their pivot point accounts for differences between the color of the stars in the observed galaxy and the color used to anchor the zero-point in their calibration. We apply this correction to F814W magnitudes in the high-fidelity photometry catalogs on a per-star basis. We apply the color-correction before fitting for the TRGB to increase the contrast of the TRGB discontinuity in the LF.

A similar metallicity effect appears for the stars near the TRGB in \jwst filters. When observed in the $F090W$ filter, the TRGB is found to be approximately constant as a function of color up to a red color limit that depends on the second filter paired with the $F090W$ filter. Stars redder than the characteristic color appear fainter, an effect dominated by higher metallicity in the atmospheres of RGB stars \citep{Anand2021,Newman2024b}. In the middle panel of \autoref{fig:ngc1380_trgb} we demonstrate the color dependent TRGB brightness in the \jwst\ $F090W$ and $F150W$ filters. \citep{Newman2024b} calibrated the \jwst\ TRGB color dependence and zeropoint for 18 combinations of NIRCam wide filters, including $F090W-F150W$ vs $F090W$. They report that the TRGB brightness is consistent with being constant in $F090W$ over the color range $1.15\text{ mag }\leq F090W-F150W\leq1.68\text{ mag}$ The stars we observe in NGC~1380 span a color range of $1.4\text{ mag }\leq F090W-F150W\leq2.8\text{ mag}$.  At present there is no empirical calibration for TRGB stars beyond the color of $F090W-F150W=1.68$ mag in the literature, though this is an active area of study. Therefore, we consider only stars for the TRGB brightness measurement in NGC~1380 within the calibrated color range from \citep{Newman2024b}.

For all CMDs, \hst\ or \jwst\, we impose color and apparent magnitude constraints to select stars that are consistent with RGB stars. Stars blueward of the RGB (e.g., main sequence or helium-burning stars) can reduce the strength of the discontinuity corresponding to the TRGB in the LF and bias the final measurement. All photometry is foreground-extinction-corrected before fitting (see column 7 in \autoref{tab:observations_summary}). 

\begin{deluxetable*}{l | lll}
    \tablewidth{0pt}
    \tablecaption{Summary of the TRGB Distance Measurements     \label{tab:trgb_stats}}
    \tablehead{
    \colhead{Galaxy} & \colhead{$m^{\text{TRGB}}_{\text{F814W}}$ (mag)} & \colhead{$\mu^{\text{TRGB}}_{\text{F814W}}$ (mag)}  & \colhead{D$^\text{TRGB}_{\text{F814W}}$ (Mpc)} 
    }
    \startdata
    NGC~1316 (SN~2006dd) & $27.45^{+0.06}_{-0.01} \text{ (stat) }$ & $31.50^{+0.06}_{-0.03} \text{ (stat) } \pm0.04 \text{ (sys) }$ & $19.95^{+0.58}_{-0.29} \text{ (stat) } \pm0.35 \text{ (sys) }$\\
    NGC~1404 (SN~2007on, SN~2011iv) & $27.18^{+0.09}_{-0.07} \text{ (stat) }$ & $31.23^{+0.09}_{-0.07} \text{ (stat) } \pm0.04 \text{ (sys) }$ & $17.62^{+0.74}_{-0.59} \text{ (stat) } \pm0.31 \text{ (sys) }$\\
     NGC~4457 (SN~2020nvb) & $27.07^{+0.01}_{-0.03} \text{ (stat) }$ & $31.12^{+0.01}_{-0.03} \text{ (stat) } \pm0.04 \text{ (sys) }$ & $16.75^{+0.08}_{-0.23} \text{ (stat) } \pm0.29 \text{ (sys) }$\\
     \hline
    \hline
    & $\quad\quad\quad\quad$ $m^{\text{TRGB}}_{\text{F090W}}$ (mag) & $\quad\quad\quad\quad\quad\quad\mu^{\text{TRGB}}_{\text{F090W}}$ (mag) & $\quad\quad\quad\quad $ D $^\text{TRGB}_{\text{F090W}}$ (Mpc) \\
    \hline
     NGC~1380 (SN~1992A) & $27.00\pm0.01 \text{ (stat) }$ & $31.32\pm0.03 \text{ (stat) } \pm0.04 \text{ (sys) }$ & $18.37\pm0.27 \text{ (stat) } \pm0.34 \text{ (sys) }$\\
     NGC~4636\tablenotemark{a}  (SN~2020ue) & $26.813\pm0.035 \text{ (stat) }$ (NRCA1)& $31.12\pm0.07$ & $16.8\pm0.6$ \\
     $\ldots$ & $26.608\pm0.035\text{ (stat) }$ (NRCA2) & $\ldots$ & $\ldots$
     \enddata
     \tablenotetext{a}{The values reported here for NGC~4636 are not measured in this study and are adopted from \citet{Anand2025}. We reproduce their results which are provided separately for NIRCam detectors A1 and A2 (NRCA1 and NRCA2, respectively). We note that \citet{Anand2025} use a different flux calibration for their \jwst\ photometry than was used in this study. The flux calibration results in a known offset in the recovered magnitudes of stars \citep{Anand2024a,Anand2024b}. The flux calibration is accounted for in the distance modulus with the proper choice of TRGB zero-point.}
     \tablecomments{We adopt a zero-point for the F814W TRGB $M^{TRGB}_{F814W} = -4.049 \pm 0.015\text{ (stat) }\pm0.035\text{ (sys) }$ mag \citep{Freedman2021}. For the JWST-based TRGB measurement in NGC~1380 we adopt the zero point $M^{TRGB}_{F090W} = -4.32\pm0.02\text{ (stat) }\pm0.04\text{ (sys) }$ mag from \citet{Newman2024b}. We report TRGB magnitudes with associated statistical uncertainties from the fitting routine only (\autoref{sec:trgb_method}). For the distance moduli ($\mu$) and distances (D), we report both uncorrelated (stat) and correlated (sys) uncertainties. Uncorrelated uncertainties include a conservative $0.5 \, A_V$ added in quadrature to the TRGB magnitude uncertainty. Correlated uncertainties include the zero-point uncertainties.}
\end{deluxetable*}

\subsection{TRGB Fitting Results}
We present our final TRGB measurements for the targets NGC~1316, NGC~1380, NGC~1404, and NGC~4457 in \autoref{tab:trgb_stats} and compare to results published in the literature. The table includes the TRGB magnitudes with statistical uncertainties, and the distance moduli and distances with separated statistical and systematic uncertainties. For galaxies observed with \hst we use the F814W TRGB zero-point from \citet{Freedman2021} of $M^{TRGB}_{F814W} = -4.049 \pm 0.015 \text{ (stat) } \pm0.035\text{ (sys) }$ mag. For NGC~1380 we use the \jwst F090W TRGB zero-point from \citet{Newman2024b} of $M^{TRGB}_{F090W} = -4.32\pm0.02\text{ (stat) } \pm0.04\text{ (sys) }$ mag. 

In the middle panels of \autoref{fig:ngc1316_trgb} through \autoref{fig:ngc4457_acs_trgb}, we present CMDs containing only high-fidelity sources for the three galaxies. The TRGB (dashed black lines) and statistical uncertainties (shaded orange regions) are marked in each CMD. We also include CMDs of only sources rejected/excluded from the high-fidelity catalogs. We include the final TRGB fits in these CMDs as a reference point. The number of rejected sources varies between the targets. 

We measure an \hst TRGB magnitude and distance modulus to NGC~1316 $m^{TRGB}_{F814W}=27.45^{+0.06}_{-0.01}\text{ (stat)}$ mag and $\mu=31.50^{+0.06}_{-0.03}\text{ (stat) }\pm0.04\text{ (sys) }$ mag. This result is in excellent agreement with the result from \citet{Hatt2018} who reported a TRGB magnitude of $m^{TRGB}_{F814W}=27.40\pm0.04$ mag and, after applying the zeropoint adopted in this study, a corresponding extinction-corrected distance modulus $\mu=31.42\pm0.04\text{ (stat) }\pm0.04\text{ (sys) }$ mag. 

For NGC~1404 we compare our \hst TRGB results of $m^{TRGB}_{F814W}=27.18^{+0.09}_{-0.07}$ mag and $\mu=31.23^{+0.09}_{-0.07}\text{ (stat) }\pm0.04\text{ (sys) }$ mag to two studies in the literature. We note that while all studies use the same \hst observations (GO-15642), their approaches to PSF-fitting photometry, TRGB fitting methodology (including the adopted zeropoint) may differ. First, we find broad agreement with \citet{Hoyt2021} who report $m^{TRGB}_{F814W}=27.32\pm0.03$ mag and extinction-corrected and zeropoint updated $\mu=31.35\pm{0.04}\text{ (stat) }\pm{0.04}\text{ (sys) }$ mag. \citet{Anand2024a} reported an \hst TRGB distance modulus of $\mu=31.28\pm{0.10}$ mag \citep[reproduced from the Extragalactic Distance Database with an updated zeropoint replacing $M^{TRGB}_{F814W}=-4.04$ mag; ][]{Anand2021} and measured a new \jwst TRGB distance modulus of $\mu=31.364\pm{0.067}$ mag. While the three reported \hst-based distance moduli show broad agreement, the best-fit distance moduli show a marginal preference towards brighter (lower/closer) values. However, including the latest \jwst-based TRGB result from \citet{Anand2024a} which skews toward a fainter magnitude (more distant) may hint at the need for additional observations of NGC~1404 that span a range of distance from the center to help reduce uncertainty.

For NGC~4457 there is only one TRGB-based distance modulus reported in the literature \citep{Li:2025}. Compared to the this study, where we find $m^{TRGB}_{F814W}=27.07^{+0.01}_{-0.03}$ mag and $\mu=31.12^{+0.01}_{-0.03}\text{ (stat) }\pm0.04\text{ (sys) }$ mag, \citet{Li:2025} report a brighter TRGB magnitude, $m^{TRGB}_{F814W}=27.00\pm0.10$ mag, and lower distance modulus, $\mu=31.0\pm0.1$ mag. Within the uncertainties we find broad agreement between the two measures. 
  
Finally, we compare to the single reported \jwst TRGB-based distance to NGC~1380 in the literature \citet{Anand2024b}. We find that our TRGB value of $m^{TRGB}_{F090W}=27.00\pm0.01\text{ (stat) }$ mag is in good agreement with the extinction-corrected TRGB magnitude reported in \citet{Anand2024a}. In particular, we find excellent agreement with their reported value measured in NIRCam module A detector 1 (A1) of $m^{TRGB}_{F090W}=27.04$ mag, while the value for NIRCam A2 is less consistent toward fainter magnitudes (i.e., NGC~1380 is further away). We use this agreement as a justification to add the galaxy NGC~4636, reported in \citet{Anand2025}, to our sample bringing the final sample size to five SN~Ia host galaxies. We adopt the distance modulus for NGC~4636 from \citet{Anand2025} of $\mu^{TRGB}_{F090W}=31.12\pm0.07$ and include the information in \autoref{tab:trgb_stats}. 

We note that all TRGB values reported in \citet{Anand2025} are based on magnitudes measured in the Vega-Vega flux calibration system, while in this study we use the current Sirius-Vega calibration. Nevertheless, the TRGB zero-points are internally consistent between the two studies.

\section{Type Ia Supernova Observations and Standardization \label{sec:snIa}}

\begin{figure}[!h]
    \centering
    \includegraphics[width=0.45\textwidth]{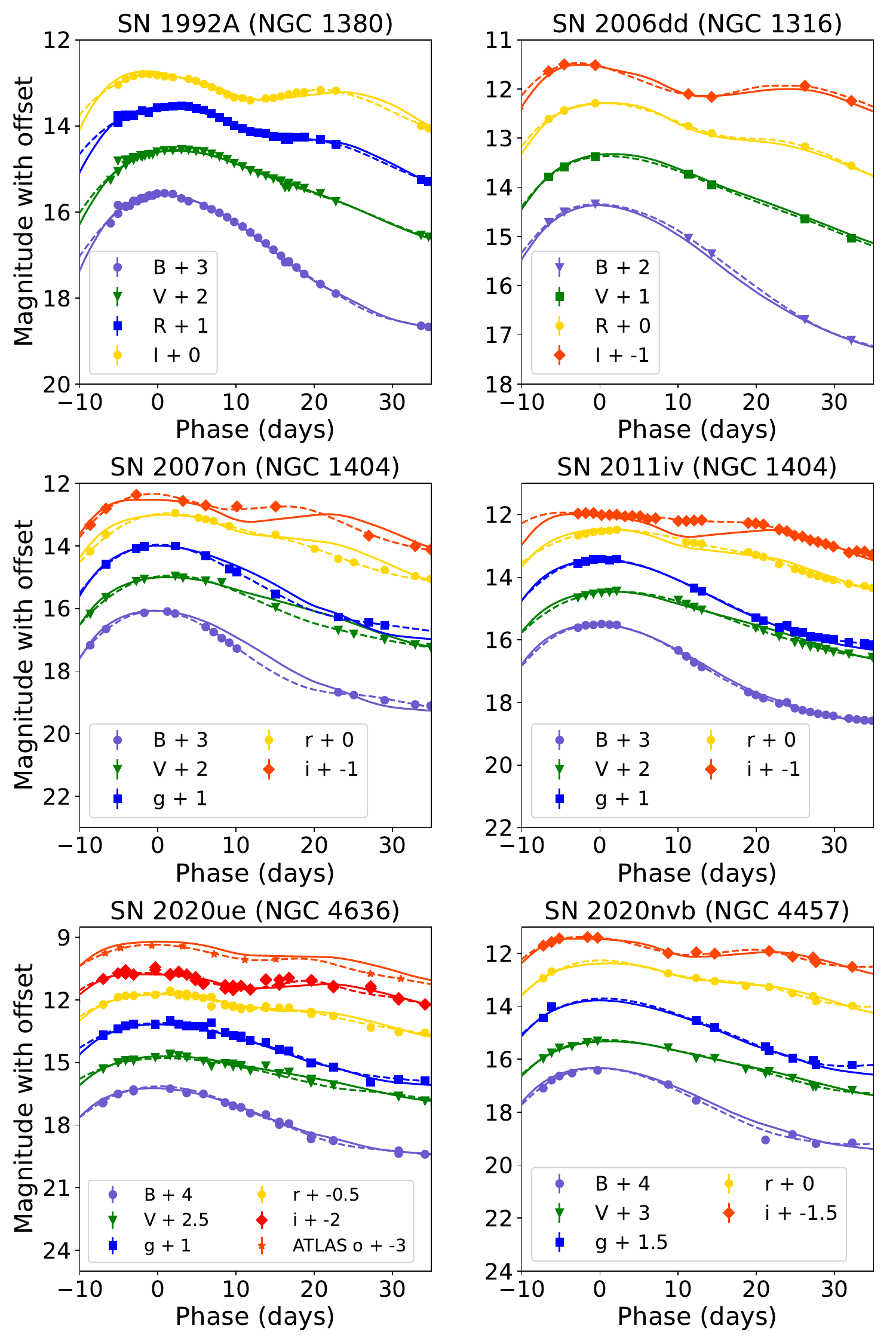}
    \caption{Photometry of the six SNe Ia for which we fit distances in this work. The epochs are presented in terms of the rest-frame phase. The solid line on each panel is the SALT2 model fit, while the dashed line is the BayeSN~model fit. BayeSN~seems to perform better for the \emph{i}-band data, potentially due to more \emph{i}-band training data and better handling of faster-evolving SNe Ia overall.}
    \label{fig:sn_comparison}
\end{figure}

The five early-type galaxies in our TRGB sample were chosen because they hosted well-measured SNe~Ia that can provide standardized distances. Three galaxies hosted four SNe~Ia with already published photometry. NGC~1316 was the host of SN~2006dd, for which we use data from \citet{Stritzinger2010}. NGC~1380 hosted SN~1992A, for which we use the photometry presented by \citet{Suntzeff:1996} and also compiled by \citet{Altavilla2004}. NGC~1404 was the host of two SNe~Ia, SN~2007on, and SN~2011iv; we use the data presented by \citet{Gall2018} to make distance measurements using these two sibling SNe~Ia. 

For the two SNe~Ia in our remaining TRGB galaxies, SN~2020ue in NGC~4636 and SN~2020nvb in NGC~4457, we present \emph{BVgri} photometry here. These data were obtained via the Sinistro cameras on Las Cumbres Observatory’s network of robotic telescopes \citep{Brown2013}, through the Global Supernova Project (GSP) collaboration. The data were reduced in the standard method with \textsc{lcogtsnpipe}\footnote{\url{https://github.com/LCOGT/lcogtsnpipe}} \citep{Valenti2016}, a PyRAF-based image reduction pipeline. Instrumental magnitudes were attained with PSF photometry after template subtraction and calibrated with external standards catalogs. Due to the brightness of both SNe and the proximity of SN~2020nvb to the center of its host, some of the observations were saturated and unusable; these epochs were removed from the analysis. The light curves of SN~2020ue and SN~2020nvb are shown in the lower panels of 
\autoref{fig:sn_comparison} and the photometry is given in \autoref{app:snphot} (\autoref{tab:2020ue_data} and \autoref{tab:2020nvb_data}). 

\begin{deluxetable*}{l | cccc | cc | c}
    \tablewidth{0pt}
    \tablecaption{SALT2 and BayeSN~Supernova Light Curve Fit Parameters and Host-galaxy Stellar Masses.  \label{tab:sn_stats}}
    \tablehead{
     \multicolumn{1}{c|}{} & \multicolumn{4}{c|}{SALT2} & \multicolumn{2}{c|}{BayeSN} & \multicolumn{1}{c}{Host Stellar Mass}\\
     \hline
    \multicolumn{1}{c|}{SN~Ia} & \colhead{$m_B$ (mag)} & \colhead{$M^i_B$ (mag)} & \colhead{$x_1$} & \multicolumn{1}{c|}{$c$} & \colhead{$A_V$ (mag)} & \multicolumn{1}{c|}{$\theta$} & \colhead{$\log M_\star/M_\odot$}     }
    \startdata
    SN~1992A   & $12.436 \pm 0.029$ & $-18.884 \pm 0.049$ & $-1.600 \pm 0.040$ & $-0.028 \pm 0.025$ & $0.06 \pm 0.04$ & $0.95 \pm 0.14$ & $10.88 \pm 0.15$ \\
    SN~2006dd  & $12.183 \pm 0.030$ & $-19.317 \pm 0.067$ & $-0.475 \pm 0.053$ & $-0.014 \pm 0.027$ & $0.15 \pm 0.06$ & $0.08 \pm 0.17$ & $11.56 \pm 0.15$ \\
    SN~2007on  & $12.922 \pm 0.023$ & $-18.308 \pm 0.093$ & $-2.061 \pm 0.043$ & $+0.022 \pm 0.022$ & $0.30 \pm 0.05$ & $1.65 \pm 0.07$ & $10.96 \pm 0.15$ \\
    SN~2011iv  & $12.348 \pm 0.023$ & $-18.882 \pm 0.093$ & $-1.745 \pm 0.033$ & $-0.024 \pm 0.021$ & $0.36 \pm 0.05$ & $1.13 \pm 0.13$ & $10.96 \pm 0.15$ \\
    SN~2020ue  & $12.024 \pm 0.022$ & $-19.096 \pm 0.073$ & $-1.839 \pm 0.031$ & $-0.025 \pm 0.019$ & $0.34 \pm 0.05$ & $1.68 \pm 0.04$ & $10.98 \pm 0.15$ \\
    SN~2020nvb & $12.140 \pm 0.026$ & $-18.980 \pm 0.048$ & $-1.419 \pm 0.065$ & $-0.055 \pm 0.024$ & $0.23 \pm 0.06$ & $1.67 \pm 0.12$ & $10.49 \pm 0.15$
    \enddata
    \tablecomments{The tabulated $M^i_B$ values are the individual SN~absolute magnitudes, obtained using just the SALT2 fit peak $m_B$ and the TRGB distance modulus, $M^i_B= m_B - \mu_{\rm TRGB}$. These are distinct from the standardized absolute magnitudes $M_B$ obtained after applying supernova corrections. SN Ia host galaxy stellar masses are based on estimates in \citet{Leroy2019} and are adjusted based on the TRGB-based distances reported in \autoref{tab:trgb_stats}.}
\end{deluxetable*}

Cosmological application of SN~Ia photometry in the optical requires standardization of their absolute magnitudes, for which many methods exist \citep[e.g.,][]{Phillips1993,Hamuy1996,Riess:1996,Perlmutter:1997,Tripp1998,Guy2005,Jha2007,Burns2011,Kenworthy2021}. Here we focus on two approaches to SN~Ia light curve standardization, SALT2 \citep{Guy2007} and BayeSN~\citep{Mandel2022}. 

SALT2 is the most commonly used methodology for SN~Ia cosmology; it fits a multicolor SN~Ia light curve with three parameters: $m_B$, which represents the apparent magnitude in the \emph{B} band\footnote{Technically, SALT2 fits in flux space using an amplitude parameter $x_0$ to scale the model flux to the observations. The model is constructed such that the peak $B$-band apparent magnitude is given by $m_B = -2.5\log x_0 + 10.5$ \citep[e.g.,][]{Taylor2023}.}; $x_1$, which parameterizes the decline rate; and $c$, which reflects the SN~color (corresponding roughly to \emph{B}$-$\emph{V} at peak). A lower $x_1$ indicates a faster-evolving light curve, and a higher \textit{c} denotes a redder color. The fiducial ``standard'' SN~Ia with $x_1 = 0$ and $c = 0$ is established during training of the model; here we adopt the SALT2 training by \citet{Taylor2023}. The light-curve shape parameter $x_1$ is further normalized so that its standard deviation over the cosmological SN~Ia sample is $\sigma(x_1) \approx 1$. The color parameter $c$ combines the effects of intrinsic variations in SN~Ia color with dust reddening in the SN host galaxy; we separately correct for Milky Way dust assuming fidelity of the \citet{Schlegel1998} dust maps and \citet{Schlafly2011} recalibration. 

SALT2 light-curve fit parameters for our TRGB-calibrated SN~Ia are given in \autoref{tab:sn_stats} and the fits are shown in \autoref{fig:sn_comparison}. We can immediately recognize that our sample is a biased draw from the SN~Ia population: all of the objects have $x_1 < 0$; indeed five of the six calibrators have $x_1 < -1.4$. A selection of early-type host galaxies with TRGB distances clearly leads to a fast-declining sample of SNe~Ia. This correlation of SN~Ia decline rate with host-galaxy environment is well known \citep[e.g.,][]{Branch:1996,Hamuy:2000,Howell:2001} and has clear import in cosmological applications (see \autoref{sec:H0}).

For the SALT2 parameterization, standardization is based on the \citet{Tripp1998} method, with linear corrections based on $x_1$ and $c$: 
\begin{equation}
\mu_\text{SALT2} = m_{B} + \alpha\,x_1 - \beta\,c - M_{B} \label{eqn:tripp}
\end{equation}
where $\mu_\text{SALT2}$ represents the inferred distance modulus, and 
$\alpha$, $\beta$, and $M_{B}$ are global parameters determined from a fit for a given sample. Depending on the cosmological application, $M_B$ is either calibrated by other distances (as we will do here with TRGB distances to infer $H_0$) or, if only relative distances are needed (e.g., in high-redshift SN~Ia cosmology), $M_B$ can be marginalized over, assuming a value of $H_0$. 

Large cosmological SN~Ia samples show distance modulus residuals that are correlated with environmental properties, e.g., the ``mass-step'' in which SN~Ia from low (stellar) mass host galaxies and high-mass host galaxies standardize to slightly different absolute magnitudes \citep{Kelly2010,Sullivan2010,Lampeitl:2010}. Precision cosmology with such samples thus often includes an additional empirical correction to \autoref{eqn:tripp} based on each SN~Ia host galaxy stellar mass. 

Importantly, because we are deliberately restricting our SN~Ia sample to those hosted in massive, early-type galaxies, we cannot reliably use previous estimates of $\alpha$, $\beta$, and the mass step that were derived from larger SN~Ia samples from a broader range of hosts. SN~Ia properties correlate with host-galaxy environment before standardization (e.g., in their light-curve shape or $x_1$ distributions) as well as after standardization (e.g., the mass step). Indeed, it has recently been shown that the best-fit $\alpha$ varies depending on the decline rates ($x_1$) of the fit SNe~Ia, with a larger $\alpha$ for faster-declining SNe~Ia \citep{Garnavich2023,Larison2024,Ginolin:2025a}. Because our sample of early-type hosts preferentially selects for such fast-declining SNe~Ia, we must take care in consistently deriving and applying the empirical light-curve corrections.

In addition, because SALT2 is trained on cosmological samples where the majority of SNe~Ia have larger $x_1$ values than our objects, it is worthwhile to explore light-curve models that may be better suited to faster-declining objects. Models with conventional decline-rate indicators like $\Delta m_{15}$ and SALT $x_1$ can run into difficulty in fitting fast-declining objects, and maximum light color \citep{Garnavich:2004} or the ``color-stretch'' $s_{BV}$ \citep{Burns:2014} may better parameterize these objects. Here, we also explore using a newer framework, BayeSN, a hierarchical Bayesian SN~Ia light curve fitting method \citep{Mandel2022}. BayeSN~separates SN~Ia colors into components based on both host galaxy dust and intrinsic SN~spectral energy distribution (SED) variations. We use the version of BayeSN~that was trained by \citet{Ward2023} and present the best-fit model parameters ($A_V$, $\theta$) in \autoref{tab:sn_stats}. The BayeSN~light curve fits to the photometry are shown in \autoref{fig:sn_comparison}; they do better at capturing the \emph{i}-band behavior of our objects. Nevertheless, to best compare with other supernova cosmology analyses, we continue to use SALT2 in our distance measurements.

\section{Measuring the Hubble Constant with a Parallel Distance Ladder: Proof of Concept \label{sec:H0}}

The SALT2 light curve fits of the TRGB-calibrated SN~Ia in massive quiescent galaxies show that these objects comprise only a subset of the total SN~Ia population. Because the statistical precision on $H_0$ measurements has been typically limited by the number of calibrator SNe~Ia, interest in a ``parallel'' distance ladder using TRGB (rather than Cepheids) has generally nonetheless included all galaxies that can be measured, to get the largest sample. This means including TRGB distances measured in the halos of star-forming SN~Ia host galaxies that also were used as Cepheid calibrators. The fraction of overlapping galaxies is relatively high: 19 out of 22  CCHP TRGB galaxies in Table 3 of \citet{Freedman2025} have Cepheid observations. Similarly, 23 out of 30 TRGB galaxies in the SH0ES compilation by \citet[Table 3]{Li:2025} have Cepheid data\footnote{In these counts, we include NGC~3627 (M66) and NGC~3368 (M96) as galaxies with Cepheid observations, even though their SNe~Ia (SN 1989B and SN 1998bu, respectively) were too reddened to be considered for the \citet{Riess2022} Cepheid sample. From the \citet{Li:2025} compilation, we also count NGC~4414 (host of SN 1974G and SN 2021J) as a Cepheid galaxy. Two additional galaxies were not counted, but may have Cepheid data: NGC~4666 (host of ASASSN-14lp), and the edge-on spiral NGC~7814 (SN~2021rhu). Including these in the count would further increase the overlap between TRGB and Cepheid hosts.}. This overlap is scientifically valuable, because it enables direct comparison of TRGB and Cepheid distances to the same galaxies \citep{Riess2024,Freedman2025}, but it also limits the independence of the two approaches, using many of the same host galaxies and the same supernovae in the second and third rungs.

In this paper we are exploring a more truly parallel distance ladder, using TRGB distances to massive, quiescent SN~Ia host galaxies only, for which Cepheid distances are not possible. Because the SNe~Ia found in these environments are a special subsample, we aim to check for systematic differences in the inferred distance scale using these objects in both the calibrator and Hubble-flow samples, rather than combining them together with SNe~Ia in star-forming hosts (whether calibrated by Cepheids or TRGB). 

One downside of this approach is clearly the limited sample size of the calibrators and potentially even Hubble flow objects. Samples grow with time, however, and in this case new large surveys like the Zwicky Transient Facility \citep[ZTF;][]{Bellm:2019a,Bellm:2019b,Graham:2019} are providing a wealth of well-measured SNe~Ia that can be used not only for better statistical precision, but perhaps more importantly, to allow subsample selection to test for systematic uncertainties. We view our analysis in this section as a ``proof of concept'' for future SN cosmology. Rather than generically treating all SNe~Ia in one group, we imagine an approach with better-matched subsamples across redshifts, e.g., calibrators and the Hubble flow, with reduced or more-tailored SN standardization corrections and different sensitivities to systematics.

Recently, \citet{Rigault:2025a} present $gri$ light curves of over 3000 SNe~Ia with $z < 0.3$ as part of ZTF SN~Ia DR2, massively increasing the low-redshift SN~Ia sample, with numerous cosmological and astrophysical applications. For our purposes, we select from the large ZTF DR2 SN~Ia sample objects that are matched to the environmental and light-curve properties of our TRGB-calibrated SN~Ia in massive, quiescent galaxies.

\subsection{Fiducial Sample Selection \label{sec:fiducial}}

Here we define ``fiducial'' sample cuts that we can apply uniformly to the calibrators and the Hubble flow, in the attempt to make these samples as similar as possible. Importantly, we defined the fiducial sample (and variants) independently of the inferred distances or the value of $H_0$; we blinded our analysis using a hidden, random offset to the TRGB distances when choosing our sample cuts, and then fixed the fiducial sample and analysis choices before unblinding. 

We begin by limiting the ZTF DR2 SN~Ia sample to what \citet{Rigault:2025a} define as the ``complete'' volume-limited sample with $z_{\rm helio} < 0.06$ \citep[see also][]{Larison2024}, comprising approximately 1000 SNe~Ia. This is particularly helpful in our application where we are preferentially selecting faster-declining, lower-luminosity SNe~Ia that are less likely to be represented at larger distances in the full ZTF DR2 SN~Ia sample.

We further restrict the sample to include only the ``cosmological'' SNe~Ia ({\tt sn\_type = snia-cosmo}). In addition to enforcing light-curve quality cuts, this selection excludes objects spectroscopically classified as peculiar. In particular, we exclude 91bg-like or 86G-like objects that are also found preferentially in the massive, quiescent galaxies \citep[e.g.,][]{Taubenberger:2008, Gallagher:2008, Panther:2019} that we are calibrating with TRGB. However, none of our calibrator SNe~Ia are explicitly spectroscopically typed as 91bg-like or 86G-like, so we match this requirement in the Hubble-flow sample. With a larger calibrator sample, it may be possible to expand our analysis to include and standardize 86G-like and 91bg-like objects \citep[e.g.,][]{Graur2024}. A small number of Hubble-flow objects (SNe 2018ccl, 2019etc, 2020nef, 2020pwn, 2020sii, 2020acua, and 2020adii) are clear outliers compared to the rest of the fiducial sample, and are also excluded from our analysis.

The ZTF SN~Ia DR2 compilation includes host-galaxy information for the sample, as well as SALT2 light curve fits. To match the calibrator galaxies, in our fiducial sample we select for massive, quiescent host galaxies with $\log (M_*/M_\odot) > 10$ and rest-frame host color $g-z > 1$ \citep{Ginolin:2025a}. Similarly, to match the calibrator SN~Ia light curve properties, we restrict our fiducial sample to objects with SALT2 parameters $-2.0 < x_1 < 0$ and $-0.2 < c < +0.1$. Finally, in our fiducial sample, we require the SNe~Ia to have well-measured spectroscopic redshifts ($\sigma_z \le 0.001$) and to be in the smooth Hubble flow ($z_{\rm cosmo} > 0.023$) in the CMB frame, after applying a flow correction\footnote{\url{https://github.com/KSaid-1/pvhub}} \citep{Carr:2022}. Even with these relatively strict cuts, the large overall ZTF cosmological sample still provides 124 Hubble-flow SNe~Ia that are matched in their properties to the calibrators.

\citet{Rigault:2025a} caution against direct cosmological use of the ZTF SN~Ia DR2 sample because of uncertain absolute photometric calibration of their forced-photometry methodology at the 0.05 mag level (though relative photometry, including across filters, is accurate to 0.01 mag). The ZTF data taken after November 2019 also suffer from a detector ``pocket effect'' causing a few-percent non-linearity in the measured flux. Both of these issues will be addressed in a future ZTF SN~Ia data release. For our ``proof-of-concept'' demonstration here, we establish the ZTF zeropoints through cross-calibration against a sample of 28 SNe~Ia measured in common with Las Cumbres Observatory $gri$ photometry (reduced as described in \autoref{sec:snIa} for SN~2020ue and SN~2020nvb). We compare individual photometric measurements as well as derived SALT2 parameters, and find consistency between both approaches. For simplicity, we apply zeropoint corrections to SALT2 $m_B$ directly. For light curves peaking before November 2019, based on the SALT2 fits to the Las Cumbres photometry, we adjust the ZTF-derived peak magnitudes as follows: $m_B = m_{B,{\rm ZTF}} + 0.024 \pm 0.035$ mag. For SNe~Ia peaking after November 2019 (with the pocket effect), we find a negligible mean offset, but increase the peak magnitude uncertainty as $m_B = m_{B,{\rm ZTF}} \pm 0.026$ mag. In addition, for these objects we correct the pocket effect on the shape of the light curves with $x_1 = x_{1,{\rm ZTF}} + 0.14$ \citep{Rigault:2025a}. We also explore the effect of not making these corrections as a sample variant in our analysis. Because our cross-calibration is based on a relatively small number of objects in common, and because changes to the ZTF zeropoint would systematically move the Hubble-flow sample relative to our calibrators, we add an overall 2.5\% systematic uncertainty in our derived \ho, corresponding to the estimated 0.05 mag zeropoint uncertainty. 

\begin{figure*}[!h]
    \centering
    \includegraphics[width=0.8\textwidth]{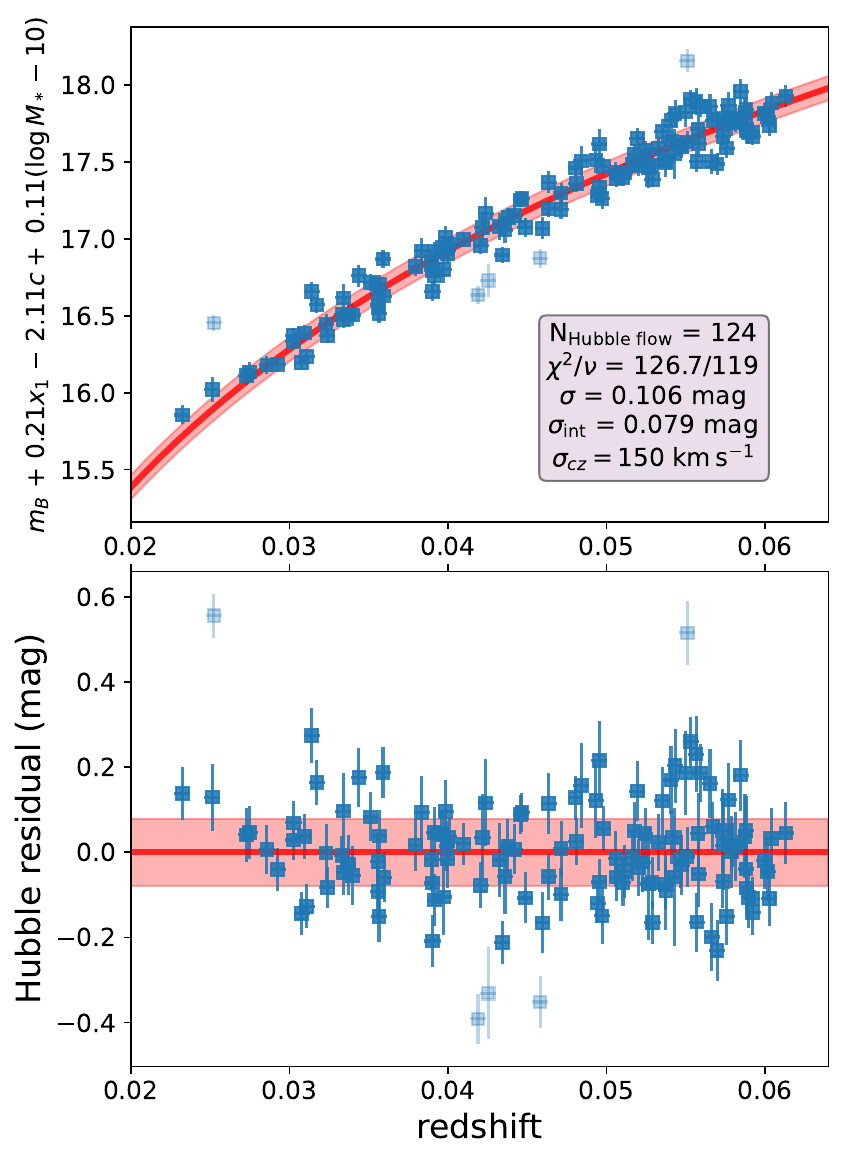}
    \caption{Hubble diagram (upper panel) and residuals (lower panel) from ZTF Hubble-flow SNe Ia \citep{Rigault:2025a} in massive, quiescent host galaxies, with our fiducial sample cuts ($0.023 < z < 0.06$, $-2 < x_1 < 0$, and $-0.2 < c < +0.1$) and self-consistent SN~Ia standardization. The small scatter (after exclusion of the outliers, shown with low opacity) validates the ZTF relative photometry and our approach of identifying a homogeneous subsample of SNe~Ia that are well matched to the TRGB calibrators with similar supernova and host-galaxy properties. }
    \label{fig:sn_hflow}
\end{figure*}

\subsection{Standardization and Model Fit \label{sec:Ia_model}}

With our fiducial sample defined and ZTF photometry corrections applied, we proceed to standardize the Hubble-flow and calibrator SNe~Ia, based on the \citet{Tripp1998} formula above, but with the inclusion of a linear host-mass correction (see below). Following the approach of \cite{Dhawan2018}, we jointly fit the TRGB-calibrated SNe~Ia that determine $M_B$ (indexed by $j$) and the Hubble-flow SNe~Ia (indexed by $k$), with
\begin{equation}\label{eqn:joint_model}
\begin{split}
\text{calibrators:} \ & M_B = m_B^j - \mu_{\rm TRGB}^j \\
& + \alpha  x_1^j - \beta  c^j + \gamma (\log M_*^j - 10) \\
\text{Hubble-flow:} \ & 5 \log d_L(H_0,\Omega_M,\Omega_\Lambda) = m_B^k - M_B \\ 
& + \alpha  x_1^k - \beta  c^k + \gamma (\log M_*^k - 10) - 25 \\
\end{split}
\end{equation}
where $M_*$ is the host-galaxy stellar mass (in solar masses), $d_L$ is the luminosity distance in Mpc, and $\Omega_M = 1 - \Omega_\Lambda = 0.3$ is assumed. The total uncertainty for each object is given by the quadrature sum of various component uncertainties and we also include covariances in the SALT2 fit parameters \citep{Marriner:2011}:
\begin{equation}\label{eqn:joint_uncertainties}
\begin{split}
\text{calibrators:} \ & \sigma_j^2 = \sigma_{m_B,j}^2 + \sigma_{\mu_{\rm TRGB},j}^2 \\
& + \alpha^2 \sigma_{x_1,j}^2 + \beta^2 \sigma_{c,j}^2 + 2\alpha \, {\rm cov}_j(m_B,x_1) \\ 
& - 2\beta \, {\rm cov}_j(m_B,c) - 2\alpha\beta \, {\rm cov}_j(x_1,c) \\
& + \gamma^2 \sigma_{{\rm log}M_*,j}^2  + \sigma_{\rm int}^2 \\
\text{Hubble-flow:} \ & \sigma_k^2 = \sigma_{m_B,k}^2 + \sigma_{z,{\rm mag},k}^2 + \sigma_{{\rm pec,mag},k}^2 \\
& + \alpha^2 \sigma_{x_1,k}^2 + \beta^2 \sigma_{c,k}^2 + 2\alpha \, {\rm cov}_k(m_B,x_1) \\
& - 2\beta \, {\rm cov}_k(m_B,c) - 2\alpha\beta \, {\rm cov}_k(x_1,c) \\
& + \gamma^2 \sigma_{{\rm log}M_*,k}^2  + \sigma_{\rm int}^2 \\
\end{split}
\end{equation}
where $\sigma_{m_B}$, $\sigma_{x_1}$, and $\sigma_c$ are the SALT2 fit uncertainties, cov() denotes their covariances, $\sigma_{\mu_{\rm TRGB}}$ is TRGB distance-modulus uncertainty (calibrators only), $\sigma_{z,{\rm mag}}$ and $\sigma_{\rm pec,mag}$ are redshift uncertainty and peculiar-velocity uncertainty (assumed to be $\pm$150 km s$^{-1}$), converted to magnitudes (Hubble-flow objects only), $\sigma_{{\rm log}M_*}$ is the logarithmic host mass uncertainty, and $\sigma_{\rm int}$ is a fit parameter for the intrinsic (unmodeled) scatter in our combined SN sample.

We use the {\tt emcee} package \cite{Foreman-Mackey2013} to perform a joint MCMC fit with six model parameters: $H_0$ (in conventional units of km s$^{-1}$ Mpc$^{-1}$), $M_B$ (mag), $\alpha$, $\beta$, $\gamma$, and $\sigma_{\rm int}$ (mag)\footnote{Because our fiducial sample restricts $-2 < x_1 < 0$, it would be better to apply the light-curve shape correction as $\alpha(x_1 + 1)$, so that there was no correction in the middle of the $x_1$ range. This would shift the definition of $M_B$ to its value for an $x_1 = -1$ supernova and help to decorrelate $M_B$ and $\alpha$ \citep{Garnavich2023}. However, to ease comparison with looser sample cuts on $x_1$ as well as with literature results, we retain the traditional centering of $M_B$ at $x_1 = 0$}. Following a convention in low-redshift supernova cosmology, we also report a version of the ``intercept of the ridge line'' that is well constrained by the Hubble-flow SNe~Ia, in magnitude units: $-5\,a_B = M_B - 5\log H_0 + 25$. Our Bayesian analysis uses uniform priors on all model parameters, enforcing $H_0 > 0$ and $\sigma_{\rm int} > 0$.

\subsection{Results \label{sec:Ia_results}}

\autoref{fig:sn_hflow} shows the high-quality Hubble diagram based on the 124 ZTF objects in the fiducial sample, after standardization, with five removed outliers (\autoref{sec:fiducial}) displayed with lower opacity. Remarkably the RMS scatter about the best-fit model is just 0.106 mag, comparable to the best results in modern supernova cosmology, with an unmodeled scatter of just $\sigma_{\rm int} = 0.079 \pm 0.009$ mag. This is a testament to the high-quality ZTF photometry \citep{Rigault:2025a} and clearly demonstrates that we have identified a homogeneous subsample of the SN~Ia population.

\begin{figure*}[!th]
    \centering
    \includegraphics[width=0.45\textwidth]{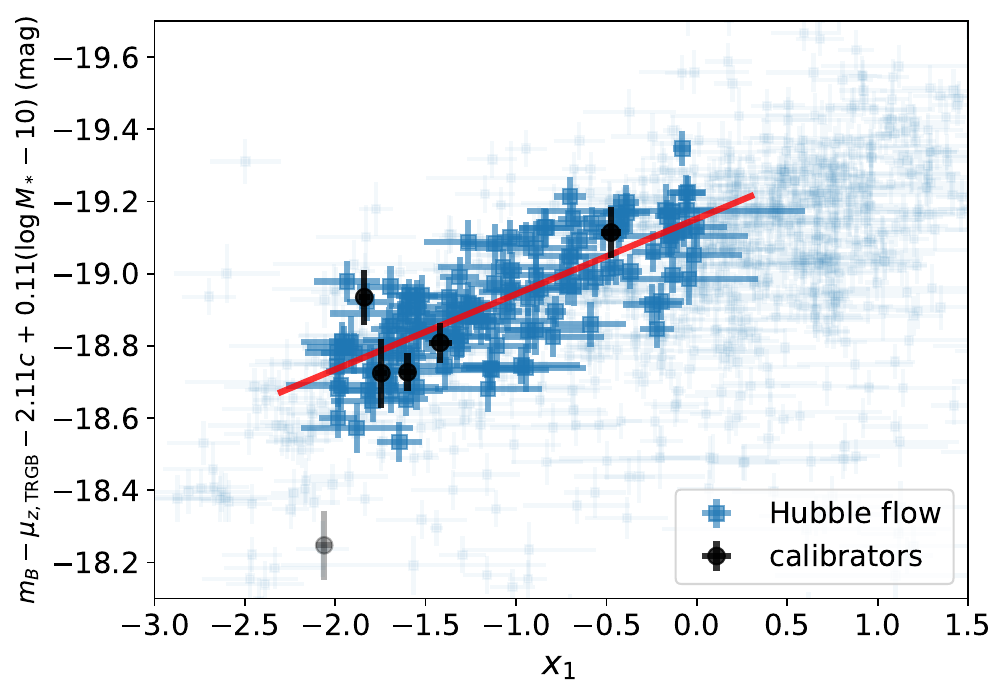}
    \includegraphics[width=0.45\textwidth]{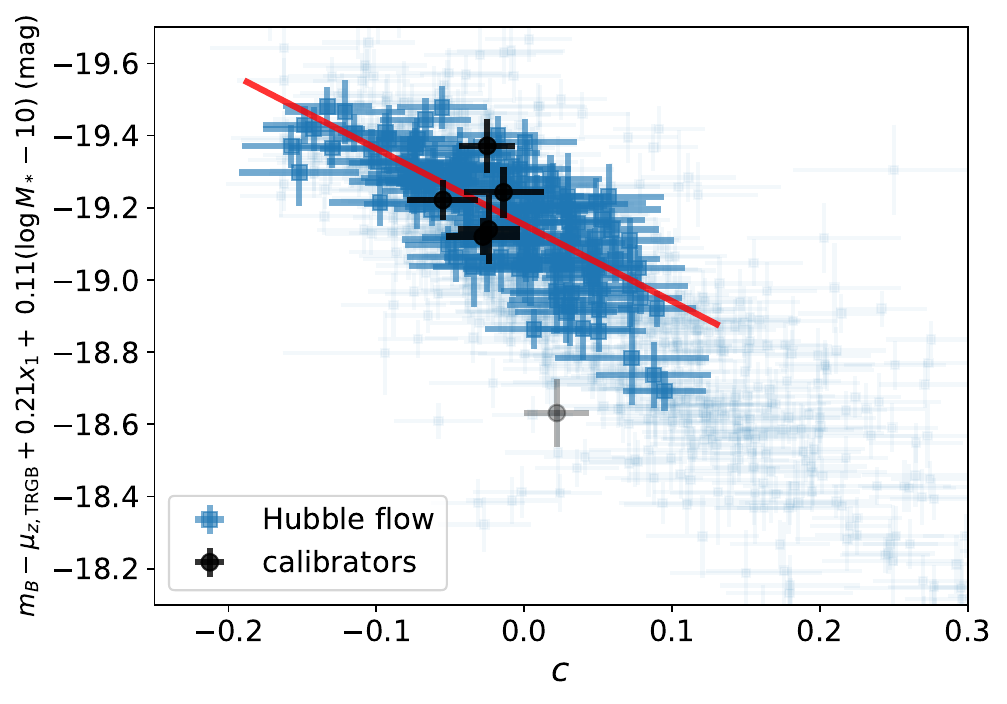}
    \includegraphics[width=0.45\textwidth]{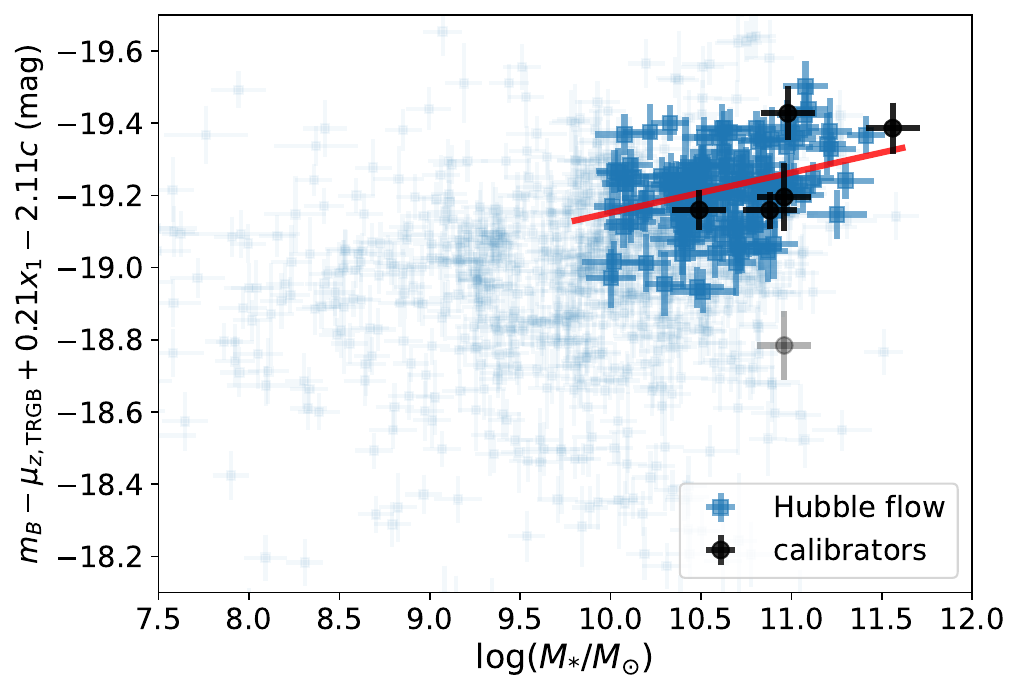}
    \caption{Standardization of Hubble flow (blue squares) and calibrator (black circles) SN~Ia samples versus SALT2 light-curve decline rate ($x_1$), SALT2 color ($c$), and host-galaxy stellar mass. The dark points comprise our fiducial sample, while the low-opacity points show the full sample. The red lines show the best-fit standardization parameters (with slopes corresponding to $\alpha$, $\beta$, and $\gamma$, respectively) from the matched, fiducial sample. The full SN~Ia population is clearly more heterogeneous and leads to significantly different estimates of the SN~Ia standardization slopes relative to our calibrator-matched fiducial subsample.} 
    \label{fig:sn_standardization}
\end{figure*}

Further validation of our approach is seen in our results for the SN correction coefficients $\alpha$, $\beta$, and $\gamma$, as illustrated in \autoref{fig:sn_standardization}. We isolate each of the correlations with light-curve shape ($x_1$), color ($c$), and host-galaxy stellar mass ($\log M_*/M_\odot$) in the figure panels. This figure also shows in low opacity the remainder of the full ZTF cosmological, volume-limited Hubble-flow sample, as well as our TRGB-calibrated SNe~Ia, demonstrating our fiducial sample cut choices to best match the calibrator and Hubble-flow objects. The data points from our fiducial sample show significantly less scatter in all panels compared to the overall population. 

Importantly, our model fit coefficients from the massive, quiescent host-galaxy SN~Ia subsample are derived for our specific sample, and differ significantly from fits to the overall SN~Ia population. Namely, we find $\alpha = 0.209 \pm 0.016$, a steeper $x_1$ correction coefficient than in typical analyses \citep[e.g., $\alpha \approx 0.15$;][]{Brout22,DESCollaboration:2024,Rubin:2025}. This is in accord with the steeper $x_1$ correction seen for fast-declining objects in other recent analyses \citep{Garnavich2023,Larison2024,Ginolin:2025a}, and can also be seen in the top left panel of \autoref{fig:sn_standardization}, where including objects with larger $x_1 > 0$ would push the best-fit slope shallower. 

Similarly, we find $\beta = 2.11 \pm 0.17$, a somewhat weaker color correction compared to full-sample analyses ($\beta \approx 3$), but consistent with SNe~Ia in luminous red galaxies at higher redshift \citep{Chen:2022}. Because the SALT2 approach uses a single color parameter that combines the effects of intrinsic color variations and host-galaxy dust reddening, it is not unexpected that our massive, quiescent host-galaxy SN~Ia subsample would show differences here \citep[likely with less dust and potentially different dust properties; e.g.,][]{Brout:2021}. The top right panel of \autoref{fig:sn_standardization} suggests that widening the color distribution of the sample would indeed favor a steeper color correction.

The lower panel of \autoref{fig:sn_standardization} explains our inclusion of a host-galaxy mass correction. Even though we have already restricted the fiducial sample to hosts with high stellar mass ($\log M_*/M_\odot > 10$), we nevertheless find a best-fit $\gamma = 0.111 \pm 0.029$ mag dex$^{-1}$, inconsistent with zero at 3.8$\sigma$ significance. Visually, a linear correction seems most appropriate for our restricted subsample, though were we to expand the sample in SN~light curve properties (i.e., wider $x_1$ or $c$ ranges) or host-galaxy stellar mass, a ``step'' correction might then be preferred. We do not speculate on the cause of this host-galaxy mass trend within our fiducial subsample; rather, we are content that in our empirical approach, its inclusion is preferred and slightly reduces the Hubble diagram scatter. We further explore this correction in analysis variants below. Understanding this issue in more detail can be complicated, based on whether host-galaxy correlations are identified after light-curve and color standardization or if they are simultaneously fit with the supernova corrections, as we do here \citep{Ginolin:2025a,Murakami:2025}.

\begin{figure*}[!th]
    \centering
    \includegraphics[width=0.9\textwidth]{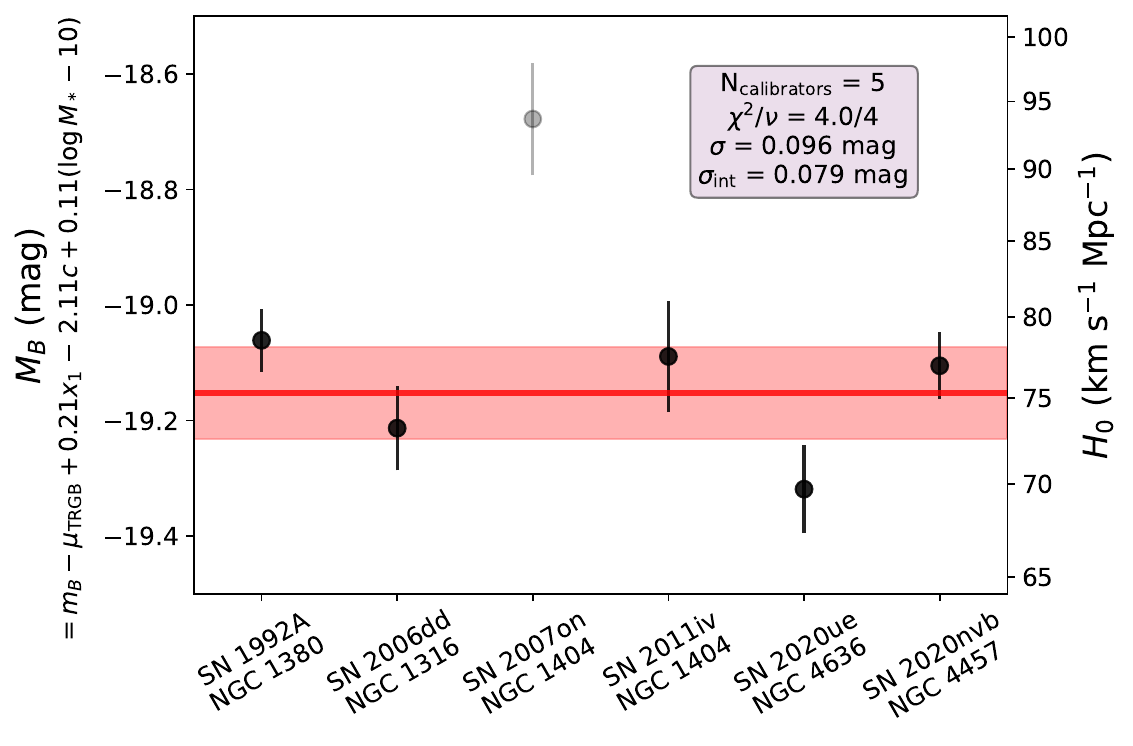}
    \caption{Standardized absolute magnitudes of the calibrator SNe~Ia based on TRGB distances presented here. The right $y$-axis shows the implications for $H_0$ assuming our fiducial Hubble-flow sample. SN~2007on in NGC 1404, with SALT2 $x_1 = -2.06 \pm 0.04$, is excluded from our fiducial sample that requires $-2 < x_1 < 0$.}
        \label{fig:sn_calibrators}
\end{figure*}

\autoref{fig:sn_standardization} highlights the comparison of the calibrator sample with the Hubble-flow objects. Choices in defining the fiducial sample cuts to best ``match'' the calibrator and Hubble-flow samples are somewhat subjective, and so we reiterate that these were made with the cosmological inferences blinded. Our fiducial cut on light-curve shape, with $-2 < x_1 < 0$ excludes the calibrator SN~2007on in NGC~1404, substantially reducing the number of fiducial calibrators from 6 to 5. We choose the threshold $x_1 > -2$ for two main reasons: first, the Hubble-flow sample becomes quite sparse below $x_1 < -2$ (\autoref{fig:sn_standardization}, top left) and second, this is also the approximate threshold where SALT2 light curve fits begin to poorly differentiate different kinds of fast-declining SNe~Ia \citep[e.g., see Figure 4 right panel of ][]{Burns:2014}. Indeed, \citet{Gall2018} note significant luminosity differences between the more extreme SN~2007on and its ``sibling'' in NGC~1404, SN~2011iv, that we retain in our calibrator sample. In our analysis, SN~2007on is more clearly an outlier compared to the other calibrators and the Hubble flow objects seen in \autoref{fig:sn_standardization}. We consider analysis variants including SN~2007on in the next section.

Our TRGB-calibrated SNe~Ia, and their individual implications for $H_0$, are shown in \autoref{fig:sn_calibrators}. With SN~2007on excluded, the scatter in the calibrators ($\sigma = 0.096$ mag) is consistent with the scatter seen in the Hubble-flow sample (\autoref{fig:sn_hflow}) and also consistent with the model (including intrinsic scatter), with the calibrators giving $\chi^2 = 4.0$ for effectively 4 degrees of freedom (5 calibrators minus 1 model parameter, $M_B$, that they constrain; the other model parameters are chiefly determined from the much larger Hubble-flow sample). 

\begin{figure*}[!h]
    \centering
    \includegraphics[width=\textwidth]{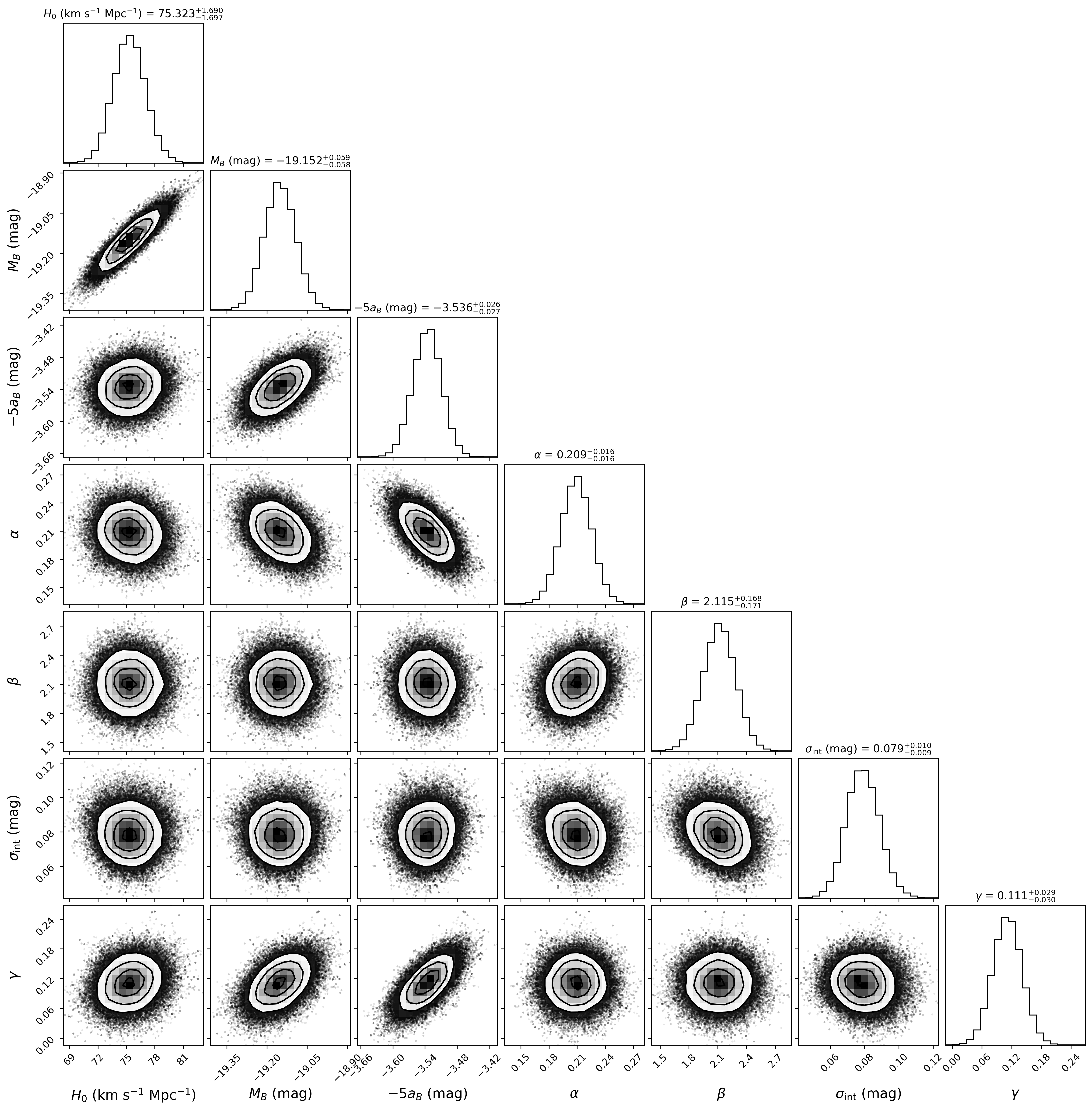}
    \caption{Corner plot visualizing the MCMC samples in our joint standardization of the fiducial ZTF Hubble-flow and calibrator samples, with the inference for $H_0$. Systematic uncertainties are not included here.}
    \label{fig:sn_corner}
\end{figure*}

A corner plot \citep{corner} of our full model fit samples is shown in \autoref{fig:sn_corner}. All model parameters are well constrained; when we quote point estimates of any of these, we use the posterior sample medians (50th percentile) with the 16th and 84th percentiles defining the approximate 1$\sigma$ confidence region. The quite symmetric posterior marginal distributions mean these estimates differ negligibly from estimates based on the sample means and standard deviations. Our full model analysis code is publicly available\footnote{\url{https://github.com/mjbnewman/ETG-TRGB-SNIa}}.

Marginalizing over all nuisance parameters, our fiducial estimate of the Hubble constant using this parallel distance ladder with TRGB-calibrated distances to fast-declining SNe~Ia, exclusively in massive, quiescent host galaxies, is $H_0 = 75.3 \pm 1.7$ (stat) $\pm$ 2.4 km s$^{-1}$ Mpc$^{-1}$. As described above, we include a 2.5\% systematic uncertainty to account for the uncertain ZTF photometric zeropoint (and subsume into this any systematic offset between the ZTF Hubble-flow light curves versus the non-ZTF calibrator light curves). We also include a 2\% (0.04 mag) correlated systematic uncertainty in the TRGB absolute calibrations and potential zeropoint offset between \hst\ F814W and \jwst\ F090W. 

Our \ho\ results are consistent with many other ``high'' values of the local expansion rate \citep[e.g.,][]{DiValentino:2025}. A simplistic comparison of our fiducial value (in this admittedly proof-of-concept approach), $H_0 = 75.3 \pm 2.9$ km s$^{-1}$ Mpc$^{-1}$ with the CMB-derived $H_0 = 67.4 \pm 0.5$ km s$^{-1}$ Mpc$^{-1}$ \citep{PlanckCollaboration:2020} corresponds to a 2.7$\sigma$ indication of the Hubble tension. We stress that our underlying data are not independent of previous TRGB+SNe~Ia analyses. In particular, we share in common with CCHP \citep{Freedman2025} the \hst TRGB data and SN light curves for NGC~1316/SN~2006dd and NGC~1404/SN~2007on+SN~2011iv, and we adopt their F814W TRGB zero point, but derive slightly different TRGB distances compared to \citet{Hoyt2021} as described above. Similarly, compared to the SH0ES analyses \citep{Riess2024,Li:2025}, we share all \hst TRGB data in common, excepting NGC~4636 \citep[adopted from][]{Anand2025}, and share SN light curves except for the SN~2020ue and SN~2020nvb photometry newly presented here. Indeed, our \hst TRGB results agree well with the SH0ES distances presented by \citet{Li:2025} and our combined TRGB+SN~inferences for \ho\ are also consistent with those shown by \citet[their Figure 3]{Li:2025} for the same objects.

\subsection{Analysis Variants \label{sec:variants}}

\begin{deluxetable*}{l|cc|cc|ccccccc}
    \tablewidth{0pt}
    \tablecaption{Model Fit Results for the Fiducial Sample and Analysis Variants. \label{tab:sn_results}}
    \tablehead{
    & \multicolumn{2}{c|}{Calibrators} & \multicolumn{2}{c|}{Hubble flow} \\
    \multicolumn{1}{l|}{Sample}  
    & \colhead{$N$}  & \multicolumn{1}{c|}{$\sigma$} 
    & \colhead{$N$}  & \multicolumn{1}{c|}{$\sigma$}
    & \colhead{$H_0$} & \colhead{$M_B$} & \colhead{$-5a_B$} & \colhead{$\sigma_{\rm int}$}
    & \colhead{$\alpha$} & \colhead{$\beta$} & \colhead{$\gamma$} \\
    & & (mag) & & (mag) & (km s$^{-1}$ Mpc$^{-1}$) & (mag) & (mag) & (mag) & & & (mag dex$^{-1}$)
    }
    \startdata
     \textbf{Fiducial:} $z>0.023$, $x_1$:[$-$2,0], $c$:[$-$0.2,+0.1]             
 & 5 & 0.096 & 124 & 0.106 & \textbf{75.3} $\pm$ \textbf{1.7} $\pm$ \textbf{2.4} & $-19.152 \pm 0.059$ & $-3.536 \pm 0.027$ & $0.079 \pm 0.009$ & $0.209 \pm 0.016$ & $2.11 \pm 0.17$ & $0.111 \pm 0.029$ \\
 Include SN~2007on, $x_1$:[$-$2.2,0]             
 & 6 & 0.192 & 134 & 0.108 & 77.0 $\pm$ 1.7 $\pm$ 2.5 & $-19.119 \pm 0.058$ & $-3.551 \pm 0.026$ & $0.083 \pm 0.010$ & $0.229 \pm 0.016$ & $2.25 \pm 0.17$ & $0.111 \pm 0.029$ \\
 Strict, fast-decliners: $x_1$:[$-$2,$-$1], $c$:[$-$0.1,0.0]               
 & 4 & 0.106 & 38  & 0.080 & 76.1 $\pm$ 1.6 $\pm$ 2.4 & $-19.253 \pm 0.088$ & $-3.659 \pm 0.078$ & $0.058 \pm 0.015$ & $0.228 \pm 0.045$ & $1.22 \pm 0.54$ & $0.040 \pm 0.045$ \\
 Loose: $z>0.015$, $x_1$:[$-$3,+1], $c$:[$-$0.3,+0.5]                 
 & 6 & 0.190 & 413 & 0.170 & 77.5 $\pm$ 2.4 $\pm$ 2.5 & $-19.001 \pm 0.071$ & $-3.448 \pm 0.016$ & $0.136 \pm 0.007$ & $0.189 \pm 0.009$ & $2.87 \pm 0.06$ & $0.150 \pm 0.026$ \\
 All Hubble-flow SNe~Ia and hosts: 
 & 6 & 0.200 & 873 & 0.211 & 78.0 $\pm$ 2.7 $\pm$ 2.5 & $-19.001 \pm 0.076$ & $-3.462 \pm 0.008$ & $0.162 \pm 0.006$ & $0.161 \pm 0.007$ & $3.05 \pm 0.04$ & $0.095 \pm 0.010$ \\
 No ZTF photometry corrections: & 5 & 0.097 & 125 & 0.104 & 76.1 $\pm$ 1.8 $\pm$ 2.4 & $-19.151 \pm 0.059$ & $-3.558 \pm 0.024$ & $0.084 \pm 0.009$ & $0.204 \pm 0.016$ & $2.08 \pm 0.16$ & $0.106 \pm 0.030$ \\
 No host stellar mass correction: & 5 & 0.117 & 124 & 0.111 & 74.0 $\pm$ 1.7 $\pm$ 2.4 & $-19.256 \pm 0.055$ & $-3.602 \pm 0.021$ & $0.087 \pm 0.009$ & $0.208 \pm 0.017$ & $2.11 \pm 0.17$ &  0 (fixed) \\
 Add SN~1994D in NGC~4526:  & 6 & 0.105 & 124 & 0.106 & 74.4 $\pm$ 1.6 $\pm$ 2.4 & $-19.179 \pm 0.056$ & $-3.537 \pm 0.026$ & $0.079 \pm 0.010$ & $0.209 \pm 0.016$ & $2.14 \pm 0.17$ & $0.110 \pm 0.029$ 
    \enddata
 \end{deluxetable*}

Though we have arrived at our fiducial sample and analysis choices blinded from the cosmological results, here we also explore plausible analysis variants for this dataset. \autoref{tab:sn_results} lists derived parameters in our joint model (with the same fitting procedure) for these variants. While there are some systematic trends based on the analysis approach, the overall cosmological inference on \ho\ does not vary greatly among these choices.

We first explore slightly expanding the SALT2 decline-rate range to $-2.2 < x_1 < 0.0$ so as to encompass the calibrator SN~2007on ($x_1 = -2.06 \pm 0.04$) that was excluded in the fiducial sample ($-2 < x_1 < 0$). This slightly increases the size of the ZTF Hubble-flow SN~Ia sample to 134 objects, with nearly identical scatter as before. However, because SN~2007on is an outlier (\autoref{fig:sn_calibrators}), the calibrator scatter increases significantly to $\sigma = 0.19$ mag and becomes somewhat inconsistent compared to the Hubble-flow scatter. Note that SN~2007on is hosted by NGC~1404, as is SN~2011iv, and the 0.4 mag luminosity difference between these sibling SNe cannot easily be reconciled. SN~2007on pulls the mean absolute magnitude scale fainter so that this analysis variant increases \ho\ by about 1.7 km s$^{-1}$ Mpc$^{-1}$ (just over 2\%) compared to our fiducial value. Compared to the other calibrators and Hubble-flow SNe~Ia in our diagnostic plots (\autoref{fig:sn_standardization}) and based on the otherwise consistent scatter in the calibrator and Hubble-flow samples, we favor continuing to identify SN~2007on as an outlier, and prefer our fiducial analysis over the variant that includes SN~2007on. A larger future calibrator sample will be needed to settle this issue more definitively.

Examining \autoref{fig:sn_standardization} further, we see that the fiducial calibrators span quite a narrow color range, and that all but one of them have $x_1 < -1$. This suggests that tighter parameter ranges might produce an even more homogeneous sample across the calibrator and Hubble-flow samples. Applying ``strict'' cuts with $-2 < x_1 < -1$, isolating the fast-declining (and perhaps ``old'') population of the bimodal $x_1$ distribution \citep[e.g.,][]{Larison2024,Ginolin:2025a}, and $-0.1 < c < 0.0$ includes 4 calibrators, but sharply reduces the Hubble flow sample to just 38 objects. Nevertheless, that severely restricted Hubble-flow sample has a residual scatter of just $\sigma = 0.080$ mag. This analysis variant slightly increases \ho\ by 1\% compared to our fiducial value. Of interest, in this variant, the host stellar mass correction slope decreases in significance to $\gamma = 0.040 \pm 0.045$ mag dex$^{-1}$. This is partly driven by the exclusion from the calibrator sample of SN~2006dd ($x_1 = -0.475 \pm 0.053$) in NGC~1316, the host galaxy with the largest stellar mass (log $M_*/M_\odot = 11.56 \pm 0.15$) and thus the largest lever arm to constrain $\gamma$. This may suggest some fragility in our determination of the host mass correction that we explore further below.

We also explore two analysis variants going in the opposite direction, loosening cuts on the Hubble-flow sample. Expanding parameter ranges to $z > 0.015$ (rather than the fiducial $z > 0.023$), with $-3 < x_1 < +1$ and $-0.3 < c < +0.5$ greatly increases the Hubble-flow sample to 413 objects, but with a large increase in Hubble residual $\sigma = 0.170$ mag. Removing constraints entirely on the ZTF Hubble-flow sample yields 873 objects with $\sigma = 0.211$ mag. As expected, loosening the sample cuts results in standardization parameters more consistent with large cosmological SN~Ia samples, i.e. with lower $\alpha$ and higher $\beta$ than in our fiducial, restricted sample. However, as seen in \autoref{fig:sn_standardization}, expanding the Hubble-flow sample creates a major mismatch when comparing to the calibrator SNe~Ia. Thus, we do not recommend adopting these variants and do not ascribe any significance to the increased \ho\ values that result. In principle, we could correct this by combining the full Hubble-flow sample with a full TRGB \citep{Li:2025} or TRGB+Cepheid \citep{Riess2024,Freedman2025} calibrator sample and accept the $\sim$0.2 mag Hubble residual scatter, but here we are advocating for the opposite approach: using smaller, more homogeneous and more precise, matched calibrator and Hubble-flow subsamples.

As described in \autoref{sec:fiducial}, our photometric recalibration of the ZTF Hubble-flow sample \citep{Rigault:2025a} is based on just 28 SNe~Ia observed in common with Las Cumbres Observatory $gri$ data. This small comparison sample may not be sufficient to adequately establish the ZTF absolute zeropoints, so we also consider an analysis variant in which no photometric corrections are applied to the ZTF data. For our fiducial sample selection, without ZTF photometric corrections, we find a 1\% increase in \ho, with otherwise negligible changes in Hubble residual or standardization parameters ($\alpha$, $\beta$, or $\gamma$). This is well within the 2.5\% systematic uncertainty we have included in our reported results, corresponding to the potential $\pm$ 0.05 mag uncertainty in the calibration offset between the calibrators and ZTF Hubble-flow sample. 

Though the host-mass correction is significantly detected in our fiducial sample ($\gamma = 0.111 \pm 0.029$ mag dex$^{-1}$), it may be largely driven by a few objects, including the calibrator SN~2006dd in the massive host NGC~1316, as described above. Thus we also consider an analysis variant removing the host-mass correction (forcing $\gamma = 0$). This results in an increase in the residual scatter for the five fiducial calibrators (from $\sigma = 0.096$ to 0.117 mag) and for the fiducial Hubble-flow sample (from $\sigma = 0.106$ to 0.111 mag) and decreases our estimate of \ho\ by 1.7\% to 74.0 $\pm$ 1.7 $\pm$ 2.4 km s$^{-1}$ Mpc$^{-1}$.

The final analysis variant we consider is to add the calibrator SN~1994D in NGC~4526. We originally excluded this object, before unblinding, because the iconic \hst image of this supernova and its host\footnote{\url{https://science.nasa.gov/asset/hubble/supernova-1994d-in-galaxy-ngc-4526/}} shows a nearly edge-on dust disk with spiral features reminiscent of a late-type galaxy. However, the \hst zoom-in on the nuclear region is misleading; NGC~4526 is classified as an S0 galaxy \citep[e.g.,][]{Burstein:1979}, has a high stellar mass with $\log M_*/M_\odot = 10.86 \pm 0.15$ \citep{Leroy2019}, and shows an integrated red color, $g - z \approx 1.6$ \citep{Kim:2014} that would have made our quiescent sample cut ($g-z > 1$). It may have been a late-type galaxy whose gas was stripped within the Virgo Cluster \citep{Young:2022}. The photometric data for SN~1994D \citep{Richmond1995} gives a SALT2 light-curve fit with $m_B = 11.704 \pm 0.029$, $x_1 = -1.558 \pm 0.026$, and $c = -0.099 \pm 0.025$, parameters that put SN~1994D squarely within the fast-declining population similar to our other calibrators (\autoref{fig:sn_standardization}). Our adherence to the blinding procedure prevents us from adding SN~1994D back to our fiducial sample, but in retrospect, it should have been included. Using its light-curve fit parameters, host stellar mass, and the reported \hst TRGB distance to NGC~4526 of $\mu_{\rm TRGB} = 31.00 \pm 0.07$ mag \citep{Hatt:2018,Freedman2019,Li:2025}, SN~1994D gives a standardized $M_B = -19.32 \pm 0.08$ mag, at the brighter end of the calibrator sample (\autoref{fig:sn_calibrators}). Our results from redoing the joint fit including SN~1994D are given in the last row of \autoref{tab:sn_results}: the average $M_B$ brightens by 0.027 mag, correspondingly lowering our estimate of the Hubble constant by 1.1\% to 74.4 $\pm$ 1.6 $\pm$ 2.4 km s$^{-1}$ Mpc$^{-1}$. The residual scatter in the calibrators increases to 0.105 mag, now nearly identical to the residual scatter in the Hubble-flow sample. We recommend including SN~1994D and NGC~4526 in any similar future analyses. 

\section{Discussion and Conclusion \label{sec:conclusion}}

The properties of SNe~Ia correlate with the properties of the environment from which they arise \citep[e.g.,][]{Hamuy1995, Sullivan2010, Lampeitl:2010, Xavier2013, Toy2023, Larison2024, Ginolin:2025a, Ginolin:2025b, Senzel:2025, Aubert2025}, and residual correlations remain even after standardization \citep{Kelly2010, Burns2018, Rigault2020, Uddin2020, Toy2025}. Thus, it is important to select a cosmological sample of SNe~Ia with consistent host and light curve properties to avoid systematic effects \citep[e.g.,][]{Wojtak:2023,Gall:2024,Wojtak:2025}. Here we have shown the impact of such a selection on an inference of $H_0$, with calibrator and Hubble flow SNe~Ia that self-consistently come from early-type, massive host galaxies and which have similar light curve properties. Our calibrator sample comprises all nearby SNe~Ia in massive, quiescent hosts for which TRGB measurements are possible. We further make use of the large sample of SNe~Ia from the ZTF SN~Ia DR2 presented by \citet{Rigault:2025a}, downselecting to a fiducial sample of 124 SNe~Ia that pass selection criteria informed by the light curve and host properties of both the Hubble-flow and calibrator samples.

We have shown that by consistently choosing only SNe~Ia in massive, quiescent hosts, we not only avoid potential systematics but in fact improve on the scatter found in other analyses, with a Hubble-flow residual RMS of only 0.106 mag, for a relatively small sample of 124 objects. This represents at least a 20\% improvement when compared to the scatter in current Hubble-flow distance ladder measurements \citep{Burns2018,Riess2022,Rigault:2025a}. Our calibrator SN~Ia sample also shows improved scatter at 0.096 mag compared to the 0.13--0.19 mag scatter seen in Cepheid or combined \emph{HST}+\emph{JWST} calibrator samples \citep{Riess2022,Freedman2025,Li:2025}, though here our calibrator sample size of 5 objects is not yet large enough to draw a robust conclusion. With our current sample, we measure $H_0 = 75.3 \pm 2.9$ km s$^{-1}$ Mpc$^{-1}$, which is offset by 2.7$\sigma$ from the CMB-derived value \citep{PlanckCollaboration:2020} but is consistent with other local-Universe measurements \citep{DiValentino:2025}. For the same analysis with the additional calibrator SN~1994D in NGC~4526, we recover $H_0 = 74.4 \pm 2.9$ km s$^{-1}$ Mpc$^{-1}$, corresponding to a 2.4$\sigma$ tension with the Planck results.

There are several points of caution for our analysis. Our selection of massive, quiescent SN~Ia host galaxies is consistent across the second and third rungs of the distance ladder, but the first rung, the geometric anchor setting the TRGB zeropoint, includes star-forming galaxies like NGC~4258 \citep{Freedman2021,Newman2024b}. If there were some unidentified systematic difference in the TRGB between the star-forming galaxy anchors and early-type galaxies, this would lead to a bias in our inferred Hubble constant. 

Moreover, in our supernova analysis we do not derive or apply ``bias corrections'' in our distances, even though we are relying on a sample of faster-declining, lower-luminosity SNe~Ia that are more susceptible to sample selection effects. We mitigate this issue by restricting our Hubble flow sample to the ZTF ``volume-limited'' sample with $z_{\rm helio} < 0.06$. Below this redshift there is no discernible trend in SN~Ia light curve parameters with distance, whereas at higher redshifts the distributions of $x_1$ and $c$ begin to show the effects of the lowest-luminosity objects disappearing from the sample \citep{Larison2024,Rigault:2025a}. Nevertheless, a full forward simulation \citep[e.g.,][]{Kessler:2017} or a more rigorous, hierarchical Bayesian model \citep[e.g.,][]{Rubin:2015,Rubin:2025,Hinton:2019} would be helpful to check this. Selection effects in our calibrator sample, heterogeneously discovered and followed-up, would be more difficult to model, though we can have confidence that contemporary surveys do not disproportionately miss even fast-declining SNe~Ia in nearby galaxies ($D \lesssim 20$ Mpc) within the reach of our TRGB distances.

Similarly, our supernova standardization approach may be problematic for our particular sample. We use SALT2 \citep{Guy2007} light curve fits with \citet{Tripp1998} standardization, the most commonly used approach in supernova cosmology today. It is especially convenient because SALT2 fits are reported in the data release \citep{Rigault:2025a} and we can easily and consistently derive these fits for the calibrator sample. However, as discussed in \autoref{sec:snIa}, SALT2 is not ideally suited for some of our faster-declining SNe~Ia (e.g., \autoref{fig:sn_comparison} shows BayeSN giving better light-curve fits, especially in $i$-band). It would be worthwhile to revisit the cosmographic analysis with alternative light-curve fitting and standardization across the calibrator and Hubble-flow samples, using tools that excel at modeling the fast-declining end of the SN~Ia population preferentially found in massive, early-type hosts.

Though we have focused on TRGB calibrations of these SNe~Ia, contrasting our host and supernova selection from a Cepheid-based distance ladder, our approach is most similar to recent work using surface-brightness fluctuation (SBF) distances to similar kinds of galaxies and hosted supernovae. In particular, \citet{Garnavich2023} construct a distance ladder and measure the Hubble constant using SNe~Ia primarily in massive, early-type galaxies, calibrated through the infrared SBF \citep{Blakeslee:2021,Jensen:2021}. Their SBF distances are cross-calibrated to an LMC Cepheid zeropoint using galaxies in the Virgo and Fornax Clusters, adding an ``extra'' rung, but otherwise their SNe~Ia analysis also focuses on the different supernova properties in early-type hosts. They also refit the SALT2 standardization for the fast-declining subsample (divided based on a cut made jointly on supernova $x_1$ and host galaxy stellar mass) with 24 SBF calibrator SNe~Ia and 175 Hubble-flow SNe~Ia with $0.02 < z < 0.25$, drawn from the Pantheon+ compilation \citep{Brout22,Scolnic2022}. \citet{Garnavich2023} identified the steeper decline-rate correlation (larger $\alpha \approx 0.23$) for the fast-declining subsample, as confirmed by \citet{Larison2024} and \citet{Ginolin:2025a} and that we see here ($\alpha \approx 0.21$). However, they find only a somewhat lower color standardization coefficient ($\beta \approx 2.7$) compared to full samples, whereas our fiducial results give $\beta = 2.1 \pm 0.2$. In addition, whereas we find a significantly lower scatter in our fiducial TRGB-calibrator and ZTF Hubble-flow SN~Ia subsamples compared to the full SN~Ia distribution, \citet{Garnavich2023} find $\sigma = 0.160$ mag for their 24 SBF-calibrator SNe~Ia and $\sigma = 0.144$ mag for their 175 Hubble-flow SNe~Ia (see their Table 3), similar to their overall SN~Ia sample. Further investigation of these differences in our results is warranted; our stricter sample cuts on the more homogeneous volume-limited (thus lower-redshift) ZTF Hubble-flow sample may be part of the explanation. \autoref{tab:sn_results} shows a trend toward lower values of $\beta$ and Hubble residual scatter as the sample cuts become more strict.

Despite these differences in detail, our Hubble constant inference is quite consistent: \citet{Garnavich2023} measure $H_0 = 74.8 \pm 2.9$ km sec$^{-1}$ Mpc$^{-1}$ for their fast-declining subsample of IR SBF calibrators and Pantheon+ Hubble-flow SNe~Ia, nearly identical to our fiducial measurement with TRGB calibrators and ZTF Hubble-flow objects. Indeed, SBF distances on their own can push out far enough into the Hubble flow to constrain \ho\ even without SNe~Ia. \citet{Blakeslee:2021} measure $H_0 = 73.3 \pm 2.5$ km sec$^{-1}$ Mpc$^{-1}$ using an SBF second and third rung and a joint Cepheid+TRGB calibration. Recently, these SBF distances have been recalibrated using \jwst TRGB observations \citep{Anand2024a,Anand2025}, yielding $H_0 = 73.8 \pm 2.4$ km sec$^{-1}$ Mpc$^{-1}$ \citep{Jensen2025}. All of these estimates, along with our fiducial measurement, show excellent consistency in an early-type galaxy distance ladder incorporating TRGB, SBF, and/or SNe~Ia.

The biggest limitation in our parallel distance ladder using only massive, quiescent galaxies is the small sample size of TRGB calibrator SNe~Ia in the second rung. Increasing this sample will help definitively establish if we have been able to identify a higher-precision subsample, as suggested by the fiducial ZTF Hubble-flow objects. \emph{JWST} observations provide a promising avenue to further expand our TRGB-calibrator sample. Its higher angular resolution and increased sensitivity is capable of measuring the TRGB in galaxies five times more distant (so a volume increased $125$ times) compared to \hst \citep[see, e.g.][]{McQuinn2019,Anand2024b,Newman2024a,Newman2024b}. The future Habitable Worlds Observatory, with an expected launch in the 2040s, holds the potential to expand the accessible range for TRGB distances beyond the Coma Cluster \citep[$D\sim100$ Mpc;][]{Anand2025b}. Such an increase in accessible volume would enable TRGB calibration of many more massive early-type SN~Ia host galaxies. TRGB calibration of nearby host galaxies of ZTF-observed SNe~Ia would be especially useful, mitigating a potential systematic uncertainty in the photometric zeropoint between the calibrators and Hubble-flow sample. In addition, TRGB-calibrated SBF distances to early-type galaxies can be extended out to $\sim$300 Mpc \citep{Cantiello2023,Jensen2025} and continue to be used either directly or as a cross-calibration for SNe~Ia to measure \ho. Finally, nearby SNe~Ia will continue to explode, and surveys like ZTF, but also ATLAS \citep{Tonry2018}, YSE \citep{Jones:2021,Aleo:2023}, and LS4 \citep{Miller2025} will be able to provide SNe~Ia for both the calibrator and Hubble-flow samples on a self-consistent photometric system.

Though we have shown that a matched calibrator and Hubble-flow sample of SNe~Ia in early-type, massive host galaxies can provide a homogeneous sample with low scatter and a precise measurement of $H_0$, it is important to note the limitations in extending this approach for cosmology at higher redshift. It is likely that the supernova sample differences we see, such as the preponderance of fast-declining, low $x_1$ objects, are driven by a physical parameter like progenitor age that evolves monotonically over the history of the Universe \citep{Childress:2014,Graur:2015,Rose:2019,Kang:2020,Nicolas:2021,Chen:2022,Lee:2022,Wiseman:2023,Larison2024,Chung:2025}. It may not be possible, thus, to create a homogeneous, matched sample of SNe~Ia in massive, early-type hosts across a wide range of redshifts. In fact, for high-redshift supernova cosmology, the opposite selection targeting similarly young SNe~Ia at all redshifts is likely preferred. In any case the general approach that we advocate could be applied to all cosmological applications of SNe~Ia: we can take best advantage of the large increase in SN~Ia sample size from current and future surveys by selecting homogeneous subsamples based on observable properties of the supernovae and their host galaxies, in principle improving both statistical and systematic uncertainties. 

\bibliographystyle{aasjournal}
\renewcommand\bibname{{References}}
\bibliography{ms.bib}

\begin{acknowledgements}
Support for program \hst-GO-16453 was provided by NASA through a grant from the Space Telescope Science Institute, which is operated by the Associations of Universities for Research in Astronomy, Incorporated, under NASA contract NAS5-26555. This research has made use of the NASA/IPAC Extragalactic Database (NED), which is funded by the National Aeronautics and Space Administration and operated by the California Institute of Technology. This work makes use of observations from the Las Cumbres Observatory network. The LCO team is supported by NSF grants AST-1911225 and AST-1911151. KAB is supported by an LSST-DA Catalyst Fellowship; this publication was thus made possible through the support of Grant 62192 from the John Templeton Foundation to LSST-DA. Support was provided by Schmidt Sciences, LLC. for MD. Supernova research at Rutgers University is supported by NSF grant AST-2407567 and DOE award DE-SC0010008.
\end{acknowledgements}

\appendix
\section{Validation of the TRGB in the NGC~4457 from the UVIS Parallel Observation}\label{sec:ngc4457_uvis_dist}
A second field in NGC~4457 was imaged with the UVIS instrument in \hst-GO-16438 via a coordinated parallel observation (see \autoref{sec:observations}). The UVIS pointing was in a region of lower crowding at a greater distance from the dense nucleus of NGC~4457 than the ACS imaging. We reduced these data with \Dolphot{} and produced a high-fidelity photometric catalog using our crowding-based spatial cut methodology (\autoref{sec:SpatialCut}). 

\begin{figure*}[!bthp]
\centering
    \includegraphics[width=0.3\textwidth, trim=0 0.4cm 0 0]{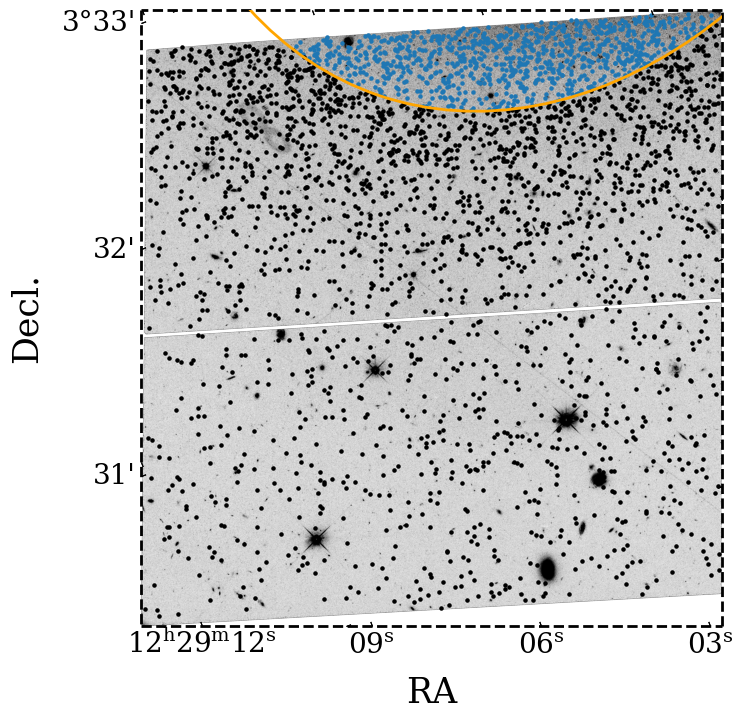}
    \includegraphics[width=0.3\textwidth]{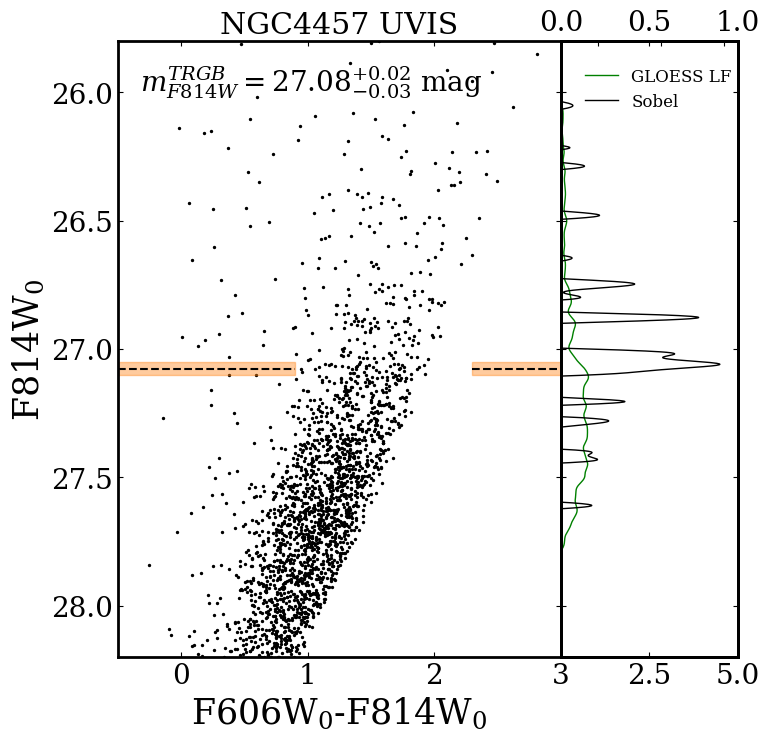}
    \includegraphics[width=0.3\textwidth]{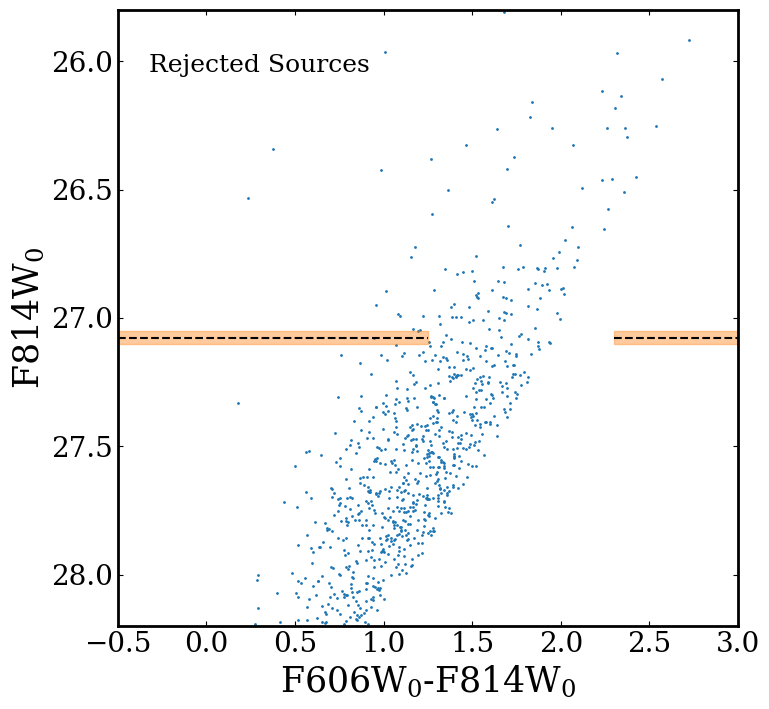}
    \caption{TRGB analysis for the parallel UVIS observations of NGC~4457. This field is located further away from the dense nucleus of NGC~4457. It is a less crowded field, but it also has far fewer sources. Left: the spatial distribution of the sources in the UVIS field. Stars removed with spatial cuts (blue points) are towards the center of NGC~4457. Middle: the high-fidelity UVIS CMD. Our best-fit TRGB is indicated (solid black line) with uncertainties (shaded orange) with a value of $m^{TRGB}_{F814W}=27.08^{+0.02}_{-0.03}$ mag. Right: CMD of sources excluded from the spatial cuts. The TRGB measured in the UVIS data is in excellent agreement with that measured from the ACS field ($27.07^{+0.01}_{-0.03}$).}
    \label{fig:ngc4457_uvis}
\end{figure*}

We then measured the TRGB magnitude using the same method for the ACS field (see \autoref{sec:trgb_method}) to provide a validation check on the ACS-based TRGB measurement. \autoref{fig:ngc4457_uvis} presents the results from our spatial selection (left) and TRGB fit (right). We determine spatial cuts following the method described in \autoref{sec:SpatialCut}. The CMD in the middle panel clearly shows a TRGB edge and is well fit by the maximum likelihood method. Notably, the TRGB apparent magnitude measured from the ACS and UVIS fields are in excellent agreement (i.e., well within $1\sigma$). The right panel shows sources exclude (blue points) from consideration.

\section{LCO photometry of SN~2020ue and SN~2020nvb \label{app:snphot}}

For the light curve fitting of SN~2020ue and SN~2020nvb, we relied on Las Cumbres Observatory (LCO) photometric data that has not yet been published. The reduction process for these data is summarized in \autoref{sec:snIa}. 

\startlongtable
\begin{deluxetable}{ccccc}
\tablecaption{LCO observations of SN~2020ue. All photometry is from template-subtracted images. \label{tab:2020ue_data}}
\tablehead{
\multicolumn{1}{c|}{Telescope} & \multicolumn{1}{c|}{MJD} & \multicolumn{1}{c|}{Filter} & \multicolumn{1}{c|}{Mag} & \multicolumn{1}{c}{Mag Err} 
}
\startdata
1m0-04 & 58861.30 & B & 14.779 & 0.022 \\
1m0-04 & 58861.30 & B & 14.789 & 0.021 \\
1m0-04 & 58861.30 & V & 14.660 & 0.024 \\
1m0-04 & 58861.30 & V & 14.648 & 0.024 \\
1m0-04 & 58861.31 & g & 14.585 & 0.008 \\
1m0-04 & 58861.31 & g & 14.571 & 0.007 \\
1m0-04 & 58861.31 & r & 14.592 & 0.008 \\
1m0-04 & 58861.31 & r & 14.593 & 0.007 \\
1m0-04 & 58861.31 & i & 14.820 & 0.007 \\
1m0-04 & 58861.32 & i & 14.839 & 0.008 \\
1m0-10 & 58862.02 & B & 14.252 & 0.022 \\
1m0-10 & 58862.02 & B & 14.239 & 0.022 \\
1m0-10 & 58862.02 & V & 14.144 & 0.024 \\
1m0-10 & 58862.02 & V & 14.141 & 0.024 \\
1m0-10 & 58862.03 & g & 14.051 & 0.008 \\
1m0-10 & 58862.03 & g & 14.035 & 0.008 \\
1m0-10 & 58862.03 & r & 14.105 & 0.009 \\
1m0-10 & 58862.03 & r & 14.103 & 0.010 \\
1m0-10 & 58862.03 & i & 14.325 & 0.008 \\
1m0-10 & 58862.04 & i & 14.328 & 0.008 \\
1m0-11 & 58862.66 & B & 13.954 & 0.024 \\
1m0-11 & 58862.66 & V & 13.948 & 0.030 \\
1m0-11 & 58862.66 & V & 18.340 & 0.155 \\
1m0-11 & 58862.66 & g & 13.654 & 0.010 \\
1m0-11 & 58862.67 & g & 13.671 & 0.009 \\
1m0-11 & 58862.67 & r & 13.816 & 0.007 \\
1m0-11 & 58862.67 & r & 13.914 & 0.008 \\
1m0-11 & 58862.67 & i & 14.013 & 0.008 \\
1m0-11 & 58862.68 & i & 14.136 & 0.010 \\
1m0-08 & 58866.38 & B & 12.890 & 0.021 \\
1m0-08 & 58866.38 & B & 12.912 & 0.021 \\
1m0-08 & 58866.39 & V & 12.795 & 0.024 \\
1m0-08 & 58866.39 & V & 12.817 & 0.024 \\
1m0-08 & 58866.39 & r & 12.715 & 0.006 \\
1m0-08 & 58866.39 & r & 12.730 & 0.007 \\
1m0-08 & 58866.39 & i & 12.999 & 0.006 \\
1m0-08 & 58866.39 & i & 12.992 & 0.006 \\
1m0-05 & 58870.28 & i & 12.800 & 0.007 \\
1m0-13 & 58874.98 & B & 12.256 & 0.022 \\
1m0-13 & 58874.98 & V & 12.088 & 0.024 \\
1m0-13 & 58874.98 & V & 12.156 & 0.024 \\
1m0-13 & 58874.99 & i & 12.839 & 0.007 \\
1m0-13 & 58874.99 & i & 12.779 & 0.011 \\
1m0-05 & 58879.24 & i & 13.235 & 0.008 \\
1m0-05 & 58879.24 & i & 13.272 & 0.008 \\
1m0-10 & 58882.03 & i & 13.490 & 0.021 \\
1m0-12 & 58883.06 & i & 13.559 & 0.017 \\
1m0-12 & 58883.06 & i & 13.339 & 0.021 \\
1m0-10 & 58884.07 & i & 13.338 & 0.014 \\
1m0-10 & 58884.07 & i & 13.353 & 0.016 \\
1m0-04 & 58885.26 & B & 13.438 & 0.022 \\
1m0-04 & 58885.26 & B & 13.443 & 0.022 \\
1m0-04 & 58885.27 & i & 13.510 & 0.007 \\
1m0-04 & 58885.27 & i & 13.528 & 0.008 \\
1m0-09 & 58887.25 & B & 13.536 & 0.026 \\
1m0-09 & 58887.25 & B & 13.511 & 0.025 \\
1m0-13 & 58888.98 & B & 13.979 & 0.024 \\
1m0-13 & 58888.98 & B & 13.851 & 0.059 \\
1m0-13 & 58888.98 & V & 13.044 & 0.028 \\
1m0-13 & 58888.98 & V & 13.039 & 0.027 \\
1m0-13 & 58888.98 & g & 13.372 & 0.010 \\
1m0-13 & 58888.98 & r & 12.874 & 0.011 \\
1m0-13 & 58888.98 & i & 13.278 & 0.014 \\
1m0-13 & 58888.98 & i & 13.032 & 0.036 \\
1m0-09 & 58890.22 & B & 13.961 & 0.025 \\
1m0-09 & 58890.22 & B & 13.950 & 0.027 \\
1m0-13 & 58893.05 & B & 14.658 & 0.029 \\
1m0-13 & 58893.05 & B & 14.667 & 0.029 \\
1m0-13 & 58893.06 & V & 13.471 & 0.032 \\
1m0-13 & 58893.06 & V & 13.459 & 0.032 \\
1m0-13 & 58893.06 & g & 13.933 & 0.012 \\
1m0-12 & 58893.10 & B & 14.544 & 0.025 \\
1m0-12 & 58893.10 & B & 14.558 & 0.024 \\
1m0-12 & 58895.92 & B & 14.736 & 0.023 \\
1m0-12 & 58895.92 & B & 14.754 & 0.024 \\
1m0-12 & 58895.93 & V & 13.574 & 0.025 \\
1m0-12 & 58895.93 & V & 13.565 & 0.025 \\
1m0-12 & 58895.93 & g & 14.227 & 0.009 \\
1m0-12 & 58895.93 & g & 14.216 & 0.010 \\
1m0-12 & 58895.93 & r & 13.259 & 0.010 \\
1m0-12 & 58895.93 & r & 13.277 & 0.013 \\
1m0-12 & 58895.93 & i & 13.401 & 0.009 \\
1m0-12 & 58895.93 & i & 13.324 & 0.014 \\
1m0-11 & 58900.71 & g & 14.959 & 0.050 \\
1m0-11 & 58900.71 & r & 13.828 & 0.016 \\
1m0-11 & 58900.71 & r & 13.829 & 0.018 \\
1m0-11 & 58900.71 & i & 13.454 & 0.050 \\
1m0-11 & 58900.71 & i & 13.345 & 0.053 \\
1m0-04 & 58904.30 & B & 15.231 & 0.025 \\
1m0-04 & 58904.30 & B & 15.368 & 0.022 \\
1m0-04 & 58904.30 & V & 14.140 & 0.032 \\
1m0-04 & 58904.30 & V & 14.103 & 0.030 \\
1m0-04 & 58904.30 & g & 14.825 & 0.009 \\
1m0-04 & 58904.30 & g & 14.837 & 0.010 \\
1m0-04 & 58904.30 & r & 14.005 & 0.009 \\
1m0-04 & 58904.30 & r & 14.049 & 0.011 \\
1m0-04 & 58904.30 & i & 13.992 & 0.017 \\
1m0-04 & 58904.30 & i & 13.915 & 0.015 \\
1m0-11 & 58907.68 & B & 15.391 & 0.022 \\
1m0-11 & 58907.68 & B & 15.406 & 0.022 \\
1m0-11 & 58907.68 & V & 14.341 & 0.025 \\
1m0-11 & 58907.68 & V & 14.336 & 0.024 \\
1m0-11 & 58907.68 & g & 14.877 & 0.009 \\
1m0-11 & 58907.68 & g & 14.884 & 0.008 \\
1m0-11 & 58907.68 & r & 14.087 & 0.009 \\
1m0-11 & 58907.68 & r & 14.070 & 0.009 \\
1m0-11 & 58907.68 & i & 14.203 & 0.008 \\
1m0-11 & 58907.69 & i & 14.231 & 0.010 \\
1m0-04 & 58910.17 & B & 15.405 & 0.023 \\
1m0-04 & 58910.17 & B & 15.437 & 0.023 \\
1m0-04 & 58910.17 & V & 14.405 & 0.025 \\
1m0-04 & 58910.17 & V & 14.440 & 0.025 \\
1m0-04 & 58910.18 & g & 14.921 & 0.009 \\
1m0-04 & 58910.18 & g & 14.894 & 0.011 \\
1m0-04 & 58910.18 & r & 14.218 & 0.010 \\
1m0-04 & 58910.18 & r & 14.208 & 0.011 \\
1m0-04 & 58910.18 & i & 14.306 & 0.018 \\
1m0-04 & 58910.18 & i & 14.264 & 0.022 \\
1m0-05 & 58913.38 & B & 15.477 & 0.022 \\
1m0-05 & 58913.38 & V & 14.524 & 0.025 \\
1m0-05 & 58913.39 & V & 14.518 & 0.025 \\
1m0-05 & 58913.39 & g & 15.013 & 0.010 \\
1m0-05 & 58913.39 & r & 14.346 & 0.010 \\
1m0-05 & 58913.39 & i & 14.554 & 0.013 \\
1m0-09 & 58916.38 & B & 15.147 & 0.025 \\
1m0-09 & 58916.38 & V & 14.214 & 0.030 \\
1m0-09 & 58916.38 & V & 14.696 & 0.027 \\
1m0-09 & 58916.38 & g & 15.238 & 0.013 \\
1m0-09 & 58916.38 & g & 15.217 & 0.013 \\
1m0-09 & 58916.38 & r & 14.358 & 0.019 \\
1m0-09 & 58916.38 & r & 14.318 & 0.031 \\
1m0-09 & 58916.38 & i & 14.305 & 0.026 \\
1m0-09 & 58916.38 & i & 14.311 & 0.015 \\
1m0-04 & 58921.14 & B & 15.524 & 0.025 \\
1m0-04 & 58921.14 & B & 15.508 & 0.024 \\
1m0-04 & 58921.14 & V & 14.679 & 0.031 \\
1m0-04 & 58921.14 & V & 14.665 & 0.028 \\
1m0-04 & 58921.14 & g & 15.136 & 0.009 \\
1m0-04 & 58921.14 & g & 15.143 & 0.010 \\
1m0-04 & 58921.14 & r & 14.634 & 0.009 \\
1m0-04 & 58921.15 & r & 14.638 & 0.009 \\
1m0-04 & 58921.15 & i & 14.751 & 0.020 \\
1m0-04 & 58921.15 & i & 14.869 & 0.010 \\    
\enddata                     
\end{deluxetable}

\startlongtable
\begin{deluxetable}{ccccc}
\tablecaption{LCO observations of SN~2020nvb. All photometry is from template-subtracted images. \label{tab:2020nvb_data}}
\tablehead{
\multicolumn{1}{c|}{Telescope} & \multicolumn{1}{c|}{MJD} & \multicolumn{1}{c|}{Filter} & \multicolumn{1}{c|}{Mag} & \multicolumn{1}{c}{Mag Err} 
}
\startdata
1m0-13 & 59031.73 & B & 13.094 & 0.020 \\
1m0-13 & 59031.73 & V & 12.977 & 0.022 \\
1m0-13 & 59031.73 & g & 12.938 & 0.006 \\
1m0-13 & 59031.74 & r & 12.942 & 0.005 \\
1m0-13 & 59031.74 & i & 13.209 & 0.008 \\
1m0-13 & 59032.79 & B & 12.791 & 0.021 \\
1m0-13 & 59032.79 & V & 12.760 & 0.024 \\
1m0-13 & 59032.79 & g & 12.517 & 0.008 \\
1m0-13 & 59032.79 & r & 12.688 & 0.010 \\
1m0-13 & 59032.79 & i & 13.073 & 0.017 \\
1m0-12 & 59033.80 & B & 12.635 & 0.020 \\
1m0-12 & 59033.80 & V & 12.574 & 0.022 \\
1m0-12 & 59033.80 & i & 12.938 & 0.006 \\
1m0-11 & 59035.41 & B & 12.522 & 0.020 \\
1m0-11 & 59035.41 & V & 12.513 & 0.022 \\
1m0-11 & 59037.40 & V & 12.364 & 0.022 \\
1m0-11 & 59037.41 & i & 12.886 & 0.005 \\
1m0-10 & 59038.70 & B & 12.417 & 0.020 \\
1m0-10 & 59038.70 & V & 12.306 & 0.022 \\
1m0-10 & 59038.71 & i & 12.910 & 0.006 \\
1m0-12 & 59047.76 & B & 12.954 & 0.020 \\
1m0-12 & 59047.76 & V & 12.560 & 0.022 \\
1m0-12 & 59047.76 & r & 12.756 & 0.005 \\
1m0-12 & 59047.76 & i & 13.492 & 0.009 \\
1m0-11 & 59051.36 & B & 13.554 & 0.021 \\
1m0-11 & 59051.36 & V & 12.957 & 0.022 \\
1m0-11 & 59051.36 & g & 13.036 & 0.006 \\
1m0-11 & 59051.36 & r & 12.934 & 0.006 \\
1m0-11 & 59051.36 & i & 13.456 & 0.010 \\
1m0-13 & 59053.73 & V & 12.956 & 0.022 \\
1m0-13 & 59053.73 & g & 13.308 & 0.006 \\
1m0-13 & 59053.73 & r & 13.052 & 0.018 \\
1m0-13 & 59053.73 & i & 13.512 & 0.007 \\
0m4-07 & 59057.71 & V & 13.368 & 0.049 \\
0m4-07 & 59057.71 & r & 13.231 & 0.062 \\
0m4-04 & 59060.25 & B & 15.047 & 0.050 \\
0m4-04 & 59060.25 & V & 13.528 & 0.030 \\
0m4-04 & 59060.25 & g & 14.015 & 0.019 \\
1m0-10 & 59060.72 & V & 13.452 & 0.022 \\
1m0-10 & 59060.72 & g & 14.166 & 0.005 \\
1m0-10 & 59060.72 & r & 13.276 & 0.005 \\
1m0-10 & 59060.72 & i & 13.416 & 0.006 \\
1m0-10 & 59063.72 & B & 14.839 & 0.021 \\
1m0-10 & 59063.72 & V & 13.710 & 0.022 \\
1m0-10 & 59063.72 & g & 14.459 & 0.006 \\
1m0-10 & 59063.72 & r & 13.508 & 0.005 \\
1m0-10 & 59063.72 & i & 13.624 & 0.008 \\
0m4-03 & 59066.35 & V & 14.024 & 0.035 \\
0m4-03 & 59066.35 & g & 14.534 & 0.028 \\
0m4-03 & 59066.35 & r & 13.611 & 0.030 \\
0m4-03 & 59066.35 & i & 13.658 & 0.041 \\
1m0-13 & 59066.72 & B & 15.192 & 0.026 \\
1m0-13 & 59066.72 & V & 14.028 & 0.025 \\
1m0-13 & 59066.72 & g & 14.704 & 0.009 \\
1m0-13 & 59066.72 & r & 13.810 & 0.011 \\
1m0-13 & 59066.72 & i & 13.833 & 0.012 \\
1m0-11 & 59071.38 & B & 15.154 & 0.024 \\
1m0-11 & 59071.38 & V & 14.167 & 0.024 \\
1m0-11 & 59071.38 & g & 14.714 & 0.010 \\
1m0-11 & 59071.38 & r & 13.978 & 0.008 \\
1m0-11 & 59071.38 & i & 14.012 & 0.012 \\    
\enddata                     
\end{deluxetable}

\end{document}